\documentclass{jfm}

\usepackage{graphicx}
\usepackage{epsfig}
\usepackage{tcolorbox}
\usepackage{amsmath}
\usepackage{amsfonts}
\usepackage{amssymb}
\usepackage{longtable}
\usepackage{dcolumn}
\usepackage{bm}
\usepackage{epstopdf}
\usepackage{color}
\usepackage{adjustbox}
\usepackage{tikz-cd}
\usepackage{hyperref}

\shorttitle{Gauge Freedom in the Morphodynamics of Fluid Deformable Surfaces}
\shortauthor{J.~Pollard, S.~C.~Al-Izzi and R.~G.~Morris}

\title{Gauge Freedom and Objective Rates in the Morphodynamics of Fluid Deformable Surfaces: the Jaumann Rate vs. the Material Derivative}
\author{
    Joseph Pollard\aff{1,2}\corresp{\email{j.pollard@unsw.edu.au}},
    Sami C.~Al-Izzi\aff{1,3}\corresp{\email{s.al-izzi@unsw.edu.au}} \and
    Richard G.~Morris\aff{1,2,3}\corresp{\email{r.g.morris@unsw.edu.au}}
    }
\affiliation{
    \aff{1}School of Physics, UNSW, Sydney, NSW 2052, Australia.
    \aff{2}EMBL Australia Node in Single Molecule Science, School of Medical Sciences, UNSW, Sydney, NSW 2052, Australia.
    \aff{3}ARC Centre of Excellence for the Mathematical Analysis of Cellular Systems, UNSW, Sydney, NSW 2052, Australia.
}

\begin{document}
\maketitle

\begin{abstract}
    Morphodynamic descriptions of fluid deformable surfaces are relevant for a range of biological and soft matter phenomena, spanning materials that can be passive or active, as well as ordered or topological.
    However, a principled, geometric formulation of the correct hydrodynamic equations has remained opaque, with objective rates proving a central, contentious issue.
    We argue that this is due to a conflation of several important notions that must be disambiguated when describing fluid deformable surfaces. These are the Eulerian and Lagrangian perspectives on fluid motion, and three different types of gauge freedom: in the ambient space; in the parameterisation of the surface, and; in the choice of frame field on the surface.
    We clarify these ideas, and show that objective rates in fluid deformable surfaces are time derivatives that are invariant under the first of these gauge freedoms, and which also preserve the structure of the ambient metric.
    The latter condition reduces a potentially infinite number of possible objective rates to only two: the material derivative and the Jaumann rate. 
    The material derivative is invariant under the Galilean group, and therefore applies to velocities, whose rate captures the conservation of momentum. The Jaumann derivative is invariant under all time-dependent isometries, and therefore applies to local order parameters, or symmetry-broken variables, such as the nematic $Q$-tensor.
    We provide examples of material and Jaumann rates in two different frame fields that are pertinent to the current applications of the fluid mechanics of deformable surfaces.
\end{abstract}


\section{Introduction}
Morphodynamic descriptions of fluid and elastic deformable surfaces have relevance across a wide variety of biological and soft matter systems, including lipid membranes~\citep{rangamani_interaction_2013, morris_mobility_2015, sahu_geometry_2020,sahu_irreversible_2017,tchoufag_absolute_2022} and vesicles~\citep{keber_topology_2014, boedec_pearling_2014, al-izzi_shear-driven_2020, al-izzi_dynamics_2020}, thin layers of cortical cytoskeleton~\citep{kruse_generic_2005, da_rocha_viscous_2022, bacher_three-dimensional_2021}, monolayers of epithelial tissue~\citep{morris_active_2019,al-izzi_active_2021, julicher_hydrodynamic_2018, khoromskaia_active_2023, hoffmann_theory_2022, blanch-mercader_integer_2021, blanch-mercader_quantifying_2021}, liquid crystal shells~\citep{khoromskaia_vortex_2017, napoli_extrinsic_2012, napoli_hydrodynamic_2016, nestler_active_2022}, actuating nematic elastomers and glasses~\citep{modes_blueprinting_2011, modes_gaussian_2011, mostajeran_frame_2017, feng_geometry_2024}, composites of microtubules and kinesin motors~\citep{sanchez_spontaneous_2012, ellis_curvature-induced_2018,pearce_orientational_2021,pearce_geometrical_2019}, and actively polymerising actin filaments~\citep{simon_actin_2019}. These examples encompass a broad range of media which can be active or passive~\citep{salbreux_mechanics_2017,al-izzi_chiral_2023}, and which often possess additional order in the form of a nematic director field. Active nematic surfaces been increasingly studied in the context of tissue mechanics and morphogenesis~\citep{salbreux_theory_2022, nestler_active_2022, khoromskaia_active_2023, rank_active_2021, vafa_active_2022,alert_fingering_2022, bell_active_2022, bacher_three-dimensional_2021}, where the nematic ordering---and in particular the topological defects---are known to play an essential role in the development of protrusions and extrusions~\citep{metselaar_topology_2019, hoffmann_theory_2022, pearce_defect-driven_2022, julicher_hydrodynamic_2018, al-izzi_active_2021,vafa_active_2022,vafa_statics_2023,hoffmann_tuneable_2023}. 

The underlying formulation of morphodynamics in this context has been explored in recent years~\citep{al-izzi_morphodynamics_2023, torres-sanchez_modelling_2019}, and while there is a growing appreciation for the importance of a principled, geometric approach to the correct hydrodynamic equations, certain aspects have remained opaque. Most notably, and independent of any particular model or material under consideration, there is still confusion about how notions that have long been well-understood in a fixed flat space translate to a moving and deforming surface, with the correct choice of objective rates being a persistently contentious issue~\citep{marsden_mathematical_1994, al-izzi_morphodynamics_2023, nitschke_observer-invariant_2022, nitschke_tensorial_2023}.

In this paper, we address the issue of objective rates in morphodynamics via three key contributions.

The first is split across Sections \ref{sec:eulerian_vs_lagrangian} and \ref{sec:gauge_freedom}. In Section \ref{sec:eulerian_vs_lagrangian} we provide a formal mathematical description of the motion of a deformable surface which clarifies the concepts of an `Eulerian' and a `Lagrangian' description. The way these terms are used in the literature is imprecise and at odds with the way they are ordinarily used in fluid dynamics in a fixed space, and is further tangled up with the question of objectivity and a gauge freedom that exists in the formulation of morphodynamics. In Section \ref{sec:gauge_freedom}, we precisely describe three group actions that relate to gauge freedoms in parts of the description of a deforming surface: a freedom in the ambient space which relates to objectivity, a freedom in the coordinate parameterisation of the surface itself, and the freedom to choose a frame field along the surface. The concepts of `Eulerian' and `Lagrangian' in the literature frequently refer to this last gauge freedom: a `Lagrangian' frame is one that moves with the fluid, while an `Eulerian' frame is one which `stays still'. A careful disambiguation of these distinct concepts provides both mathematical and physical insight. 

Secondly, in Section \ref{sec:objective_rates} we carefully examine the concept of an objective rate for a deformable surface. There are infinitely many possible choices, and no clear way to differentiate between them or otherwise select the `correct' one. We clarify the formulation of objectivity by presenting it in terms of invariance under certain gauge transformations. We also stress that it is important to impose another physically-motivated requirement on our rates, besides just objectivity: that the rate does not advect the Euclidean metric in the ambient space. With this additional requirement, the material and Jaumann rates emerge as the correct choices for the objective rate. These are shown to be invariant under different group actions. The former applies to velocities, whose rate is intimately linked with momentum conservation, and thus is synonymous with the Galilean group of transformations. The latter applies to local, symmetry-broken variables, such as the nematic $Q$-tensor, that are manifestly invariant under all time-dependent isometries. More concretely, this comes down to whether the rate of a given quantity should depend on (global) angular momentum, and therefore whether an observer is spinning: the material derivative does not account for this, whereas the Jaumann rate is `corotational'.

Our third contribution is to provide formulas for the material and Jaumann rates of various quantities involved in surface dynamics. In Section \ref{sec:formulas} we give expressions for these quantities in two different choices of frame. The first is a local coordinate chart which is not advected by the flow---an `Eulerian' parameterisation~\citep{torres-sanchez_modelling_2019,torres-sanchez_approximation_2020,sahu_arbitrary_2020}---and the second is an orthonormal frame. Detailed computations of this formulae using two different approaches---the method of moving frames, as well as the more familiar method using Christoffel symbols---are presented in Appendices \ref{appA} and \ref{appB}. We also include Appendix \ref{appC}, which overviews the geometric notions of pushforward and pullback that we make use of throughout the text.

\section{Eulerian vs. Lagrangian Perspectives}
\label{sec:eulerian_vs_lagrangian}
Consider the standard distinction between the Eulerian and Lagrangian perspectives on fluid flow inside a fixed manifold $S$. In the Eulerian perspective, we imagine standing still at a fixed point in $S$ and watching the fluid flow past us. The Eulerian specification of the flow field at a point $p \in S$ and at time $t$ is therefore a vector ${\bf v}_t(p)$ in the tangent space $\text{T}_p S$ to the point $p$ that gives the direction at which matter is flowing past us. Globally, this results in a time-dependent vector field ${\bf v}_t$ on $S$, a path in the space $\mathfrak{X}(S)$ of vector fields on $S$. By contrast, the Lagrangian perspective instead considers the medium to be made up of fluid particles, whose paths we follow through time. At time $t$, the fluid particle that was initially at point $p \in S$ has moved to some new point $\psi_t(p) \in S$. For a smooth fluid motion, this gives rise to a time-dependent diffeomorphism $\psi_t$ of $S$, a path in the diffeomorphism group $\text{Diff}(S)$. Crucially, these two perspectives are completely equivalent and can be mapped onto one another: the Eulerian flow field ${\bf v}_t$ and the Lagrangian diffeomorphism $\psi_t$ are related by
\begin{equation} \label{eq:flow1}
    {\bf v}_t \circ \psi_t = \partial_t \psi_t,
\end{equation}
where $\partial_t$ denotes partial differentiation with respect to time.

The deeper mathematical relationship between Eulerian and Lagrangian perspectives has been studied using the language of differential geometry and Lie group theory, a set of ideas originally developed by~\cite{arnold_vladimir_2014} and described in detail in the textbook of ~\cite{arnold_topological_2021}. We will not give a detailed discussion of this theory here, and instead only concern ourselves with aspects salient to the differences between ordinary fluid dynamics and the motion of deformable surfaces. Chief amongst these is the use of (\ref{eq:flow1}) to interpret $\partial_t \psi_t$ as a vector field on $S$: while this is fine in the setting of ordinary fluid dynamics, trying to carry this reasoning over to the setting of deformable surfaces requires care.

Properly, a Lagrangian motion is a path through the diffeomorphism group $\text{Diff}(S)$ (or the group $\text{SDiff}(S)$ of volume-preserving diffeomorphisms for an incompressible flow). At each time $t$ the tangent direction $\partial_t \psi_t$ to this path lies in the tangent space $\text{T}_{\psi_t} \text{Diff}(S)$ to the diffeomorphism group at the point $\psi_t$. At any point the tangent space to the diffeomorphism group can be identified with its Lie algebra, and this is nothing more than the space $\mathfrak{X}(S)$ of vector fields on $S$---thus, for any fixed $t$, $\partial_t \psi_t$ can be understood as a vector field on $S$. When considering the motion and deformation of fluid surfaces, however, this identification is more complicated, and correspondingly the relationship between the Eulerian and Lagrangian perspectives is more complicated.

Notably, when dealing with deformable surfaces there is not one manifold, but three: an abstract body $B$, which in our case is a 2D manifold because we are considering a deformable surface; an ambient space in which the motion happens, which for us will always be 3D Euclidean space $\mathbb{R}^3$ equipped with the Euclidean metric $e$; and the image $M$ of $B$ in $\mathbb{R}^3$, which is the physical surface which we observe. The ambient space is the natural analog of the fixed manifold in ordinary fluid dynamics, and must remain invariant under any dynamics. The body is not usually equipped with a metric, but often has a volume form $\mu$ which is interpreted as a density measure for an unstrained configuration, and is necessary to describe conservation of mass and incompressibility. 


To clarify, we consider a configuration of the system as an embedding $r : B \to \mathbb{R}^3$ of the body into the ambient space, whose image is the material, a smooth submanifold $M$ of $\mathbb{R}^3$. The collection of all such embeddings is a space $\text{Emb}$ which has the structure of an infinite-dimensional manifold. A motion of the system is then a path $r_t : B \to \mathbb{R}^3$ of embeddings with images $M_t$, that is, a path in $\text{Emb}$---the role played by this space is therefore analogous to the role played by the diffeomorphism group $\text{Diff}(S)$ in the case of ordinary fluid dynamics in a fixed space $S$. 

The derivative $\partial_t r_t$ defines a tangent vector to this path in the space $\text{Emb}$. Define ${\bf U}_t$ to be the value of this derivative at time $t$, an element of $\text{T}_{r_t}\text{Emb}$: this is a map ${\bf U}_t : B \to \text{T}\mathbb{R}^3$. For a deformable surface this map is the analogue of the derivative $\partial_t \psi_t$ of the diffeomorphism giving the Lagrangian description of the motion. However, note that this map is \textit{not} a vector field, either on $B$ or in $\mathbb{R}^3$, and it cannot be identified with one via the relationship \eqref{eq:flow1} used in ordinary fluid dynamics because the tangent space to $\text{Emb}$ is not isomorphic to a space of vector fields. This is one of the key differences when considering the motion of a deformable surface.

\begin{figure}
 \centering
 \includegraphics[width=\linewidth]{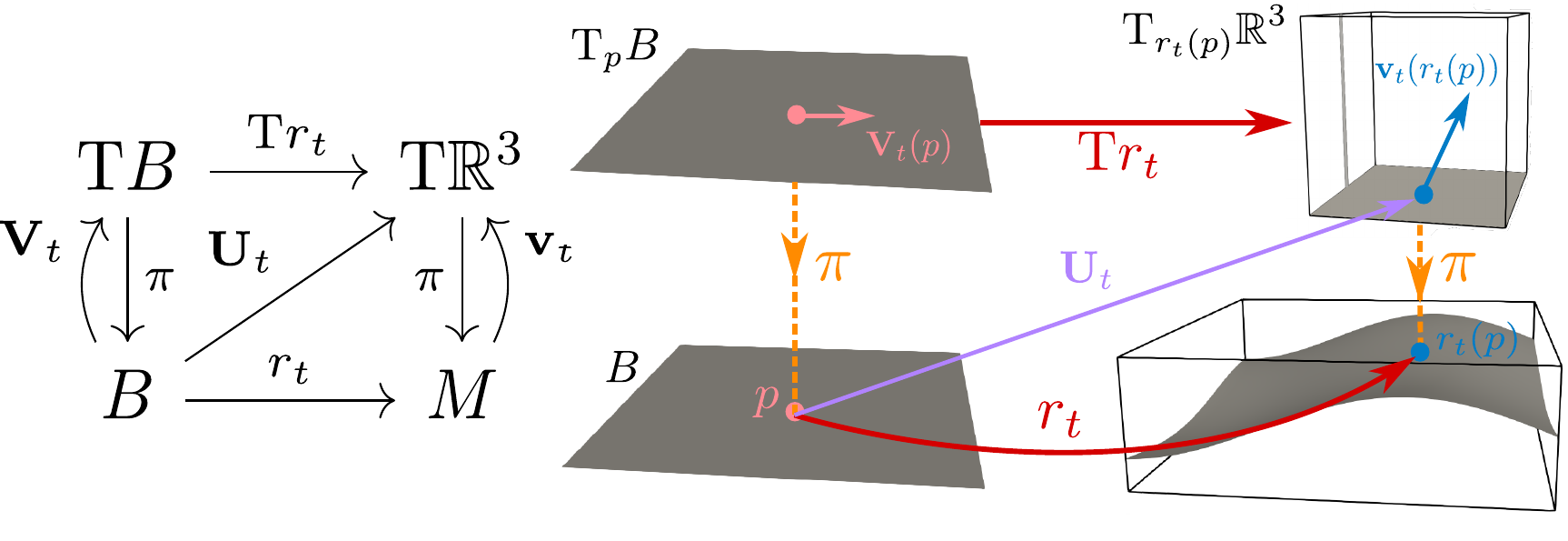}
 \caption{An embedding $r_t$ of an abstract space $B$ into $\mathbb{R}^3$ induces a vector field ${\bf V}_t$ on $B$ called the left Eulerian velocity, as well as a map ${\bf v}_t : M_t \to \text{T}\mathbb{R}^3$ on the image $M_t$ of the embedding. We interpret that latter as a vector field along $M_t$ which has a part tangent to $M_t$ but may also have a part normal to $M_t$, and refer to it as the right Eulerian velocity. Pulling back this vector field along the embedding $r_t$ `forgets' the normal part, yielding the left Eulerian velocity ${\bf V}_t$. The mathematical relationships between these maps are shown in the diagram on the left, while a more visual representation is shown on the right. Here, $\text{T}r_t$ denotes the tangent map (matrix of partial derivatives) induced by $r_t$, $\pi$ is the projection from the tangent bundle to the underlying manifold, and ${\bf U}_t = \partial_t r_t$ as described in the text.}
 \label{fig1}
\end{figure}

We can derive two vectors fields from the map ${\bf U}_t$ according to the diagram shown in Fig~\ref{fig1}. The first of the two vector fields is the `right Eulerian velocity', a map ${\bf v}_t : M_t \to \text{T}\mathbb{R}^3$ defined by,
\begin{equation}
    {\bf v}_t :=  {\bf U}_t \circ r^{-1}_t.
\end{equation}
This is a vector field in $\mathbb{R}^3$ defined along $M_t$. As such, it decomposes as ${\bf v}_t = {\bf v}^\perp_t + v^n_t{\bf n}_t$, where ${\bf v}^\perp_t$ is tangent to $M_t$ and ${\bf n}_t$ is the normal. There is also the `left Eulerian velocity', a vector field ${\bf V}_t : B \to \text{T}B$ on $B$ defined by
\begin{equation}
    {\bf V}_t := (\text{T}r_t)^{-1} \circ {\bf U}_t.
\end{equation}
Here, $\text{T}r_t$ denotes the tangent map induced by the embedding; see Appendix \ref{appC} for the definition. The right Eulerian velocity pulls back to the left Eulerian velocity, $r^*_t {\bf v}_t = {\bf V}_t$, but for purely dimensional reasons the normal component is lost and the pushforward of the left Eulerian velocity is accordingly $r_{t*} {\bf V}_t = {\bf v}^\perp_t$, the tangent part ${\bf v}_t$.

Neither of these vector fields quite corresponds to our intuitive idea of Eulerian motion: the right Eulerian velocity ${\bf v}_t$ is defined on a surface that is intrinsically moving; the left Eulerian velocity does not encode the normal motion of the surface. To properly specify the Eulerian and Lagrangian descriptions of surface motion, we need to introduce a small amount of additional structure which formally captures some intuitive notions about surface dynamics that would be unnecessary if $B$ were itself a 3D body.   

\begin{figure}
 \centering
 \includegraphics[width=\linewidth]{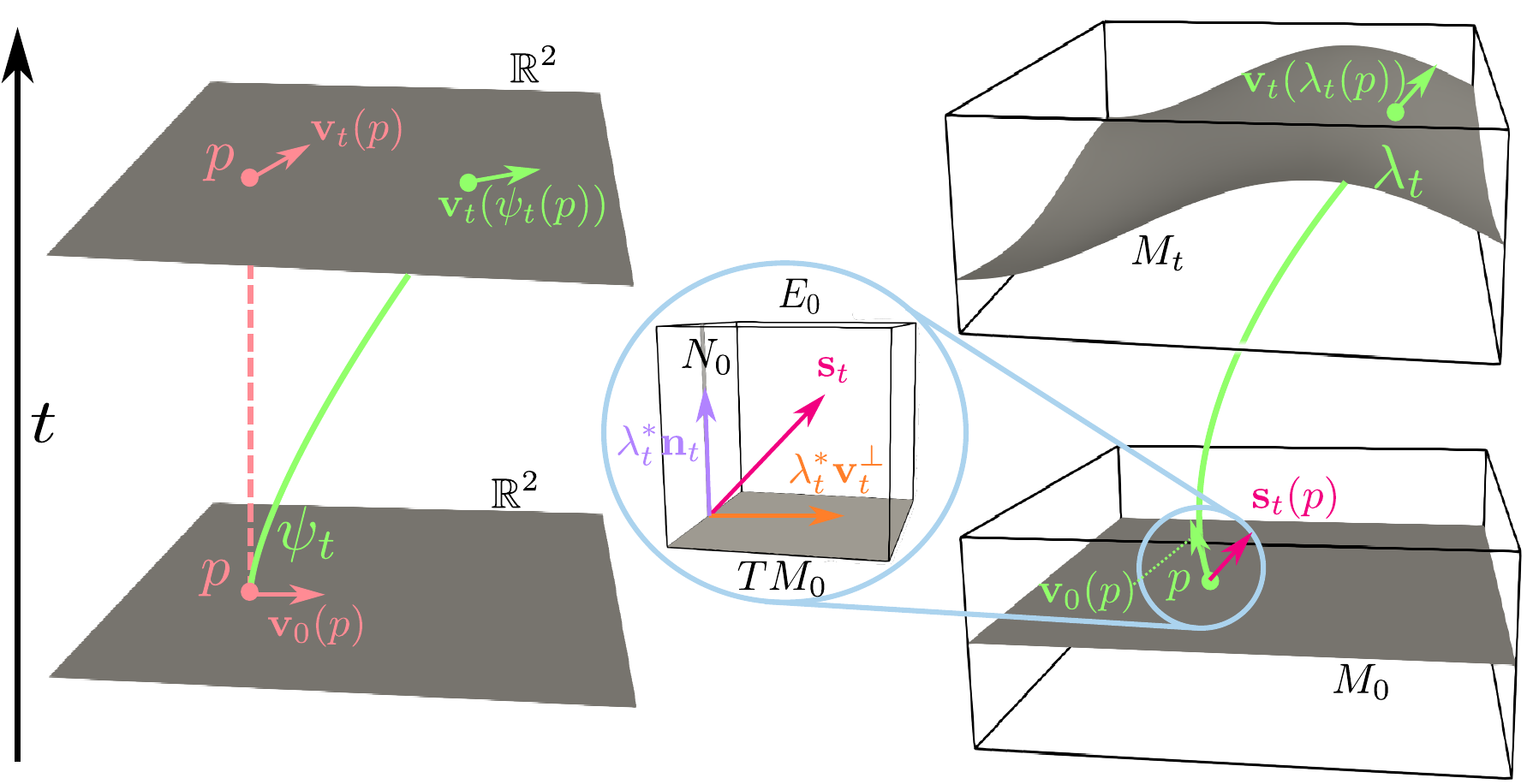}
 \caption{Comparison of the Eulerian and Lagrangian pictures for motion in a fixed surface (left) and for a deformable surface (right). In a fixed plane $\mathbb{R}^2$, the Lagrangian picture involves a diffeomorphism $\psi_t$ of $\mathbb{R}^2$ that moves a fluid particle initially the point $p$ to the point $\psi_t(p)$. We may also consider an Eulerian picture, where we stand still at the point $p$ and watch the fluid flow past us, its direction at time $t$ being given by the vector ${\bf v}_t(p)$. A deformable surface $M_t$ embedded in $\mathbb{R}^3$ undergoes a Lagrangian motion described by a diffeomorphism $\lambda_t : M_0 \to M_t$ that maps the initial surface onto the time $t$ surface. The associated right Eulerian velocity field ${\bf v}_t$ lies in the extended tangent space $E_t$ to $M_t$, as described in the text. By pulling back the entire tangent space via $\lambda_t$ (inset) as described in the text we can define a time-varying vector field ${\bf s}_t$ along $M_0$ which plays the role of the Eulerian velocity at a fixed point $p \in M_0$.}
 \label{fig2}
\end{figure}

The manifold $M_t$ has two natural vector bundles associated to it. The first is $\text{T}M_t$, its usual two-dimensional tangent space. However, we also need to consider directions that contain a component normal to $M_t$, and these lie in a three-dimensional bundle that we denote by $E_t$, the restriction of $\text{T}\mathbb{R}^3$ to $M_t$. The Euclidean metric defines an orthogonal splitting $E_t = \text{T}M_t \oplus N_t$, where $N_t$ is the one-dimensional normal bundle. The orthogonal projection of the Euclidean metric onto $\text{T}M_t$ is then exactly the induced metric on $M_t$ (the first fundamental form). The right Eulerian velocity ${\bf v}_t$ is a section of the bundle $E_t$; to be explicit, a vector field defined along $M_t$ that has both tangent and normal components. This bundle is visualised in Fig.~\ref{fig2}.

We fix an initial embedding $r_0$ with image $M_0$. The motion $r_t$ then induces diffeomorphisms $\lambda_t : M_0 \to M_t$ with $\lambda_t = r_t \circ r^{-1}_0$. The path $\lambda_t(p)$ of some point $p \in M_0$ is then naturally seen as the motion of a fluid particle in the ambient space $\mathbb{R}^3$, as thus is the a Lagrangian description of the motion. We have the relationship
\begin{equation} \label{eq:flow2}
    {\bf v}_t \circ \lambda_t = \partial_t \lambda_t,
\end{equation}
and hence the right Eulerian velocity ${\bf v}_t$ is the vector field naturally associated to the Lagrangian motion $\lambda_t$.

The Eulerian description must involve standing at a fixed point on the initial manifold $M_0$. If $M_t$ were a 3D manifold, there we would be no problem with pulling back the velocity field ${\bf v}_t$ on $M_t$ along $\lambda_t$ to give a vector field on $M_0$ that naturally corresponds to the Eulerian picture of the motion. Because we consider a 2D surface this doesn't work out for dimensional reasons, but this is a purely technical issue that can easily be resolved with a simple definition.

The map $\text{T}\lambda_t : \text{T} M_0 \to \text{T} M_t$ associates a tangent vector on $M_0$ to a tangent vector on $M_t$. Let us define a map $A_t : E_0 \to E_t$ that acts on the spaces of 3D vector fields along $M_0$ and associates them to 3D vector fields along $M_t$. To define this map, we use the splitting $E_t = \text{T}M_t \oplus N_t$ into a tangent space and a normal bundle. By linearity, it suffices for us to define that $A_t$ acts on the $\text{T}M_0$ factor exactly as $\text{T}\lambda_t$, and acts on the $N_0$ factor by mapping the unit normal ${\bf n}_0$ to $M_0$ to the unit normal ${\bf n}_t$ to $M_t$. Now, we may naturally define a section ${\bf s}_t$ of $E_0$ by
\begin{equation} \label{eq:eulerian_for_surface}
    {\bf s}_t(p) := A_t^{-1} \circ {\bf v}_t \circ \lambda_t(p). 
\end{equation}
Concretely, if ${\bf v}_t \circ \lambda_t = {\bf v}^\perp_t + v^n_t {\bf n}_t$, then 
\begin{equation}
    {\bf s}_t(p) = \lambda^*_t{\bf v}^\perp_t + (\lambda_t^* v^n_t){\bf n}_0.
\end{equation}
Then ${\bf s}_t$ is always a 3D vector field along $M_0$, as we may consider it to be the Eulerian flow field of the motion. It of course satisfies ${\bf s}_0 = {\bf v}_0$, because the map $A_0$ is just the identity map. We visualise this pullback process in Fig.~\ref{fig2}. We could also describe this in a convective representation on the manifold $B$, by pulling the whole bundle $E_t$ back along the embedding $r_t$, and taking the pullback $r^*_t {\bf s}_t$ as the left Eulerian velocity, a section of $r^*_t E_t$.

The ways in which the terms `Eulerian' and `Lagrangian' are often used in the literature---especially in computational studies~\cite{torres-sanchez_modelling_2019, torres-sanchez_approximation_2020, sahu_arbitrary_2020}---refers to something quite different from the Eulerian-Lagrangian split we have just outlined. Rather, they describe different ways of parameterising quantities on the surface in terms of a surface frame field. The freedom to choose the frame is a kind of `gauge freedom', and the physics is agnostic to the particular choice of gauge. There are, in fact, several gauge freedoms that arise in the mathematical formulation of morphodynamics, some of which leave the physics invariant and some of which don't, and these choices play an essential role in the formulation of objectivity and observer motion. We describe this in detail in the next section.

\section{Gauge Freedom}
\label{sec:gauge_freedom}
We describe gauge freedoms in the usual language of gauge theory in physics, in terms of the action of a symmetry group and the principal bundle which it defines~\citep{frankel_geometry_2011, naber_topology_2011, naber_topology_2011-1}. In order to help fix ideas, we briefly comment on how gauge freedoms manifest in the motion of a fluid in a fixed space $S$. There is a natural action of the group $\text{Diff}(S)$ on this space. It has a subgroup $\text{Isom}(S)$ of isometries, those diffeomorphisms $\phi$ that fix the metric $g$ on $S$, $\phi^* g = g$. In $n$-dimensional Euclidean space this is the Euclidean group $\text{E}(n)$ of rotations and translations. If we add in time-dependent isometries with a constant velocity, so-called Galilean boosts, then we obtain the Galilean group, $\text{Gal}$. Hydrodynamics is manifestly not invariant under the action of $\text{Diff}(S)$, but it is required to be invariant under $\text{Gal}$. This then corresponds to a gauge freedom in how we specify our equations of motion, as we are free to describe the physics itself in any inertial reference frame. Informally, objectivity is the invariance of our equations of motion under the action of the Galilean group, which can be interpreted as their invariance under a motion of a hypothetical observer in an inertial reference frame.

There is a second gauge freedom which is not concerned with the physics, but simply the representation of physical quantities. A frame field on $S$ is a choice of basis for the tangent space $\text{T}_p S$ at every point $p$, which varies smoothly on space. We should be careful to disambiguate this from the notion of an inertial reference frame or a hypothetical observer, and hence from the notion of objectivity---it is a fundamentally different concept. To describe a physical quantity that is a vector or a tensor---for example the velocity field ${\bf v}$---we pick some frame field ${\bf e}_j$---for example a coordinate frame---which spans the tangent space to $S$, and then we may write ${\bf v} = v^j{\bf e}_j$ for some set of functions $v^j$. However, the frame field is entirely arbitrary and bears no relation to the physics. We are free to choose a different frame field $\bar{\bf e}_j$ and instead write ${\bf v} = \bar{v}^j\bar{\bf e}_j$. Of course, the vector field ${\bf v}$ does not change under a change of frame, and it does not matter whether the frame field is a coordinate frame, whether it is orthonormal, or whether it varies in time. This freedom to choose the frame field is then associated with the action of $\text{Diff}(S)$ not on the manifold itself, but on the frame bundle $\text{F}S \to S$, an action which sends a frame ${\bf e}_j$ to the new frame $\psi_* {\bf e}_j$. 

Now we return to the setting of deformable surfaces. In this problem there are really {\it three} distinct gauge freedoms, and part of the confusion around objectivity and `Eulerian' vs `Lagrangian' motions has to do with a failure to properly disambiguate between them. The state of the deforming surface is captured by an element $r \in \text{Emb}$. There are two group actions on this space: $\text{E}(3)$ and $\text{Diff}(B)$ act on $\text{Emb}$ respectively from the right and the left. 

The first group $\text{E}(3)$ is the isometry group of the ambient space, and it reflects our ability to change reference frames in the ambient space (not on the surface). The second group $\text{Diff}(B)$ is the diffeomorphism group of $B$, and its action on $\text{Emb}$ reflects our ability to make arbitrary changes of coordinate system in the base space. This is a gauge freedom related to the parameterisation of the motion of the surface itself, independently on any physics or quantities defined on the surface. The third gauge freedom again concerns an action of $\text{Diff}(B)$, but this time on the frame bundle $\text{F}B \to B$; equivalently, an action of the diffeomorphism group $\text{Diff}(M_t)$ of the embedded surface at time $t$ on its own frame bundle $\text{F}M_t \to M_t$. As with motion in a fixed space, this describes our freedom in choosing a local frame field with which to represent quantities defined along the surface. 

We describe each of these group actions and their associated gauge freedoms in more detail in the following subsections.

\subsection{Gauge Freedom in the Ambient Space}
First we describe the action of the isometry group. For an isometry $\phi \in \text{E}(3)$ and embedding $r \in \text{Emb}$, the group action sends $r$ to $\phi \circ r$ and sends the image $M$ of $r$ to a different submanifold $\phi(M)$. While these two submanifolds will in general be different, they are related by a rigid body motion and not by a deformation (that is, a `pure motion') and the first and second fundamental forms induced on $\phi(M)$ are the same as those on $M$---more precisely, the pullbacks of these quantities to $B$ are equal. 

Group actions on a space define quotient spaces and principal bundles over those spaces~\citep{frankel_geometry_2011, naber_topology_2011, naber_topology_2011-1}. We define $\text{Iso}$ to be the quotient space defined by the action of $\text{E}(3)$ on $\text{Emb}$. This is the space of deformable surfaces up to isometry, which can be interpreted as surfaces with a fixed centre of mass (we can always use a translation to move this to the origin) and a fixed orientation at the centre of mass (a global rotation ensures this can always point along the $z$-axis). Alternatively, we may view it as the space of deformable surfaces with a single fixed point $p$ at which the normal direction never changes. Associated to $\text{Iso}$ is the fiber bundle,
\begin{equation} \label{eq:fibration1}
    \text{E}(3) \to \text{Emb} \to \text{Iso}.
\end{equation}
Sitting above a point in $\text{Iso}$ is the group $\text{E}(3)$ which parameterises all possible placements and orientations for the centre of mass (equivalently a fixed point $p$) of the surface---this is the gauge freedom, and fixing a given isometry returns us to a point in the full space $\text{Emb}$. A path of distinct embeddings which correspond to the same point in $\text{Iso}$ can be distinguished by an observer stood at a fixed point in space, but if the observer is allowed to move with the surface they can can change their position so that their view of it never changes. More succinctly, in $\text{Iso}$ we see only deformation and not the motion.

The action of $E(3)$ is then associated with a gauge freedom---the equations of motion are invariant under the action of $E(3)$, and so the physics does not really `see' motion in $\text{Emb}$, it only sees motion in the quotient space $\text{Iso}$. 

As we saw in the case of a fixed space, the definition of objectivity requires extending this group action to the larger space of time-dependent isometries, of which Gal is a subgroup. We discuss this, including some of the subtleties, in more detail in the next section.

\subsection{Gauge Freedom on the Deforming Surface}
The group $\text{Diff}(B)$ acts on $\text{Emb}$ as follows. Let $r \in \text{Emb}$ be an embedding and $\psi$ a diffeomorphism (coordinate change) of $B$. The action of $\text{Diff}(B)$ sends $r$ to $r \circ \psi$. The image of $B$ under $r \circ \psi$ is the same as its image under $r$, i.e., these two different embeddings define exactly the same submanifold $M$ of $\mathbb{R}^3$, but they correspond to distinct points in $\text{Emb}$. This coordinate change therefore induces no motion and no deformation of the surface. Instead of considering motion relative to $B$ we may instead consider it relative to $M_0$, the initial material surface, in which case we consider the action of the group $\text{Diff}(M_0)$---clearly isomorphic to the group $\text{Diff}(B)$---acting to change coordinates in the initial surface, with corresponding action on the diffeomorphism $\lambda_t: M_0 \to M_t$. 

The quotient space defined by the action of $\text{Diff}(B)$ on $\text{Emb}$ is the `space of membranes' $\text{Memb}$, which can alternatively be described as the collection of all images of embeddings
\begin{equation}
    \text{Memb} = \{r(B) \ | \ r \in \text{Emb} \}.
\end{equation}
From the perspective of an outside observer it is impossible to distinguish points in $\text{Emb}$ that correspond to the same point of $\text{Memb}$, even if the observer moves around. We then have an associated fiber bundle,
\begin{equation} \label{eq:fibration2}
    \text{Diff}(B) \to \text{Emb} \to \text{Memb}.
\end{equation}
Sitting above each point in $\text{Memb}$---which is nothing more than some submanifold $M$ of Euclidean space with the same topology as $B$---is a copy of the group $\text{Diff}(B)$, which can be thought of as parameterising all possible coordinate systems on $M$. 

We want to consider the effects of changing this particular gauge on our description of the motion. Let $r_t : B \to \mathbb{R}^3$ be any path of embeddings describing a motion, with image $M_t$, and let $\psi_t$ be an arbitrary time-dependent diffeomorphism of $B$. We associate to $\psi_t$ its `drive velocity' ${\bf w}^\psi_t = (\partial_t \psi_t) \circ \psi^{-1}_t$, which is a vector field on $B$ which can be seen as the velocity of an observer moving around on $B$ (not in the ambient space) starting at an initial point $p$ whose own (Lagrangian) motion is $\psi_t(p)$. A time-dependent drive velocity generates a `fictitious flow' $\partial_t {\bf w}^\psi_t$ on the surface, and thus the hydrodynamics of any physical quantity defined on the surface is not invariant under this group action even though the motion of the surface itself is. 

Concretely, if we define a new embedding $r^\psi_t := r_t \circ \psi_t$, then this embedding has the same image $M_t$ as $r_t$ and the associated right Eulerian velocity field is
\begin{equation} \label{eq:gauge_transform_velocity}
    {\bf v}_t^\psi = {\bf v}_t + r_{t*}{\bf w}^\psi_t,
\end{equation}
where ${\bf v}_t = (\partial_t r_t) \circ r^{-1}_t$ is the Eulerian velocity field associated with $r_t$. We note that since ${\bf w}^\psi_t$ is a vector field on $B$ its pushforward $r_{t*}{\bf w}^\psi_t$ is a tangent vector field on $M_t$ with no component along the normal direction. In particular, this suggests a natural choice of gauge transformation in which $\psi_t$ is chosen so that $-r_{t*}{\bf w}^\psi_t$ is exactly the tangential part of ${\bf v}_t$, and hence the velocity field in this gauge is directed along the normal direction to $M_t$. Indeed, this gauge can be defined as one in which the pullback of ${\bf v}_t^\psi$ vanishes, which obviously requires that 
\begin{equation}
    0 = r^*_t {\bf v}_t^\psi = {\bf V}_t + {\bf w}^\psi_t.
\end{equation}
We call this the `normal gauge'. This tells us that the space $\text{Memb}$ only `sees' the normal part of the motion and never the tangential part, which can always be canceled out by a relabelling of the fluid particles. Computationally, when describing the evolution of the deforming surface itself (but not quantities on the surface) it is convenient to work in the space $\text{Memb}$, moving the mesh points purely along the normal direction~\cite{torres-sanchez_modelling_2019}. Because the surface is invariant under these gauge transformations this presents no issue. However, $\text{Memb}$ ignores all evolution on the surface, and therefore we cannot describe the evolution of other quantities on the surface in this gauge, only the motion of the surface itself.

\begin{figure}
 \centering
 \includegraphics[width=0.9\linewidth]{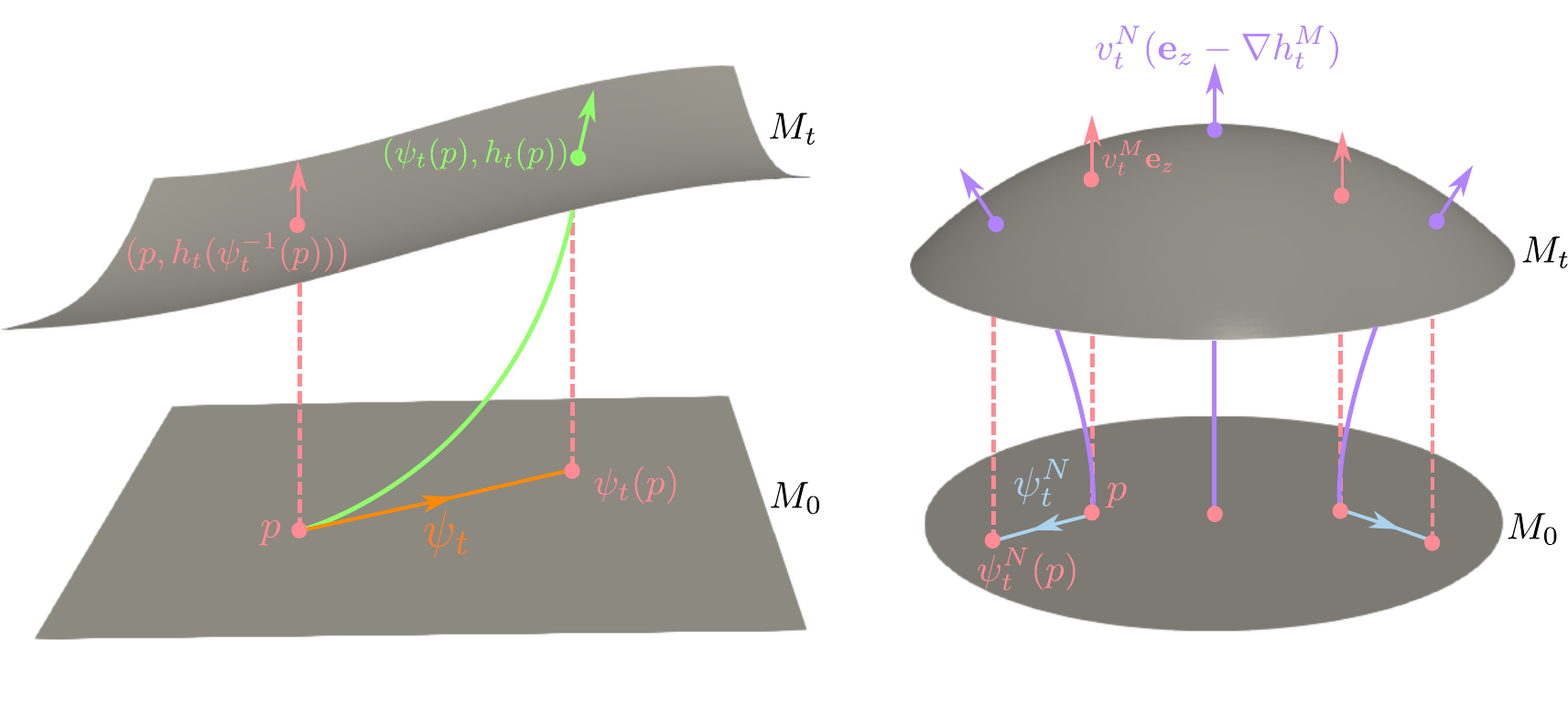}
 \caption{(Left) Construction of a Monge gauge. The evolution of an initially-flat surface (green) can be decomposed as a pair $(\psi_t(p), h_t(p))$ where $\psi_t$ (orange) is a diffeomorphism is the initial surface $M_0$ and the $h_t$ is a height function. By making a time-dependent change of gauge using the inverse $\psi^{-1}_t$ as described in the text, we can ensure the evolution is determined purely by a gauge-transformed height function $h_t \circ \psi^{-1}_t$, which ensures a fluid particle initially at the point $p$ evolves purely in the vertical direction (pink). (Right) Transition from the Monge gauge (pink) to the normal gauge (purple). This involves a diffeomorphism $\psi_t^N$ (blue) whose drive velocity is $-\nabla h^M_t$, where $h^M_t = h_t \circ \psi_t^{-1}$ is the height function in the Monge gauge. In the Monge gauge the velocity vector of the surface is ${\bf v}^M = v^M {\bf e}_z$, while in the normal gauge it is ${\bf v}^N = v^N ({\bf e}_z -\nabla h^M)$.}
 \label{fig3}
\end{figure}

For a concrete example of this gauge transformation, choose $B = \mathbb{R}^2$ to be a plane. Let us fix the usual $x,y,z$ on $\mathbb{R}^3$, and identify $B$ with the plane $M_0$ with coordinates $(x,y,0)$. Thus the deforming surface $M_t$ at time $t$ can be related to its initial value by the Lagrangian motion $\lambda_t$, which can be written in the form
\begin{equation} \label{eq:generic_motion}
    \lambda_t(x, y) = (\psi_t(x, y), h_t(x, y)).
\end{equation}
In this expression $\psi_t : \mathbb{R}^2\to \mathbb{R}^2$ is a diffeomorphism of $M_0$ and $h_t : \mathbb{R}^2 \to \mathbb{R}$ is a `height' function. At a point $p = (x,y,z)$ that lies on $M_t$ the velocity field is 
\begin{equation}
    {\bf v}_t(p) = (\partial_t \psi_t) \circ \psi_t^{-1}(x,y) + h_t(\psi^{-1}_t(x,y)){\bf e}_z,
\end{equation}
where ${\bf e}_z$ is the unit vector along the Cartesian $z$ direction in $\mathbb{R}^3$. Now we make a change of gauge using the diffeomorphism $\psi_t^{-1} \in \text{Diff}(M_0)$. The Lagrangian motion in this new gauge is $\lambda^M_t = \lambda_t \circ \psi_t^{-1}$, or in coordinates,
\begin{equation}
    \lambda^M_t(x, y) = (x, y, h_t(\psi_t^{-1}(x, y))).
\end{equation}
Defining $h^M_t = h_t \circ \psi_t^{-1}$, we see that that the motion in this gauge can be described entirely in terms of a height function. We call this the `Monge gauge'~\citep{monge_application_1850} because it is related to a local parameterisation of the surface in terms of Monge patches, and we illustrate this transition in the leftmost panel of Fig.~\ref{fig3}. 

It is easy to see the transformation from the Monge gauge to the normal gauge. The normal direction on the surface is ${\bf e}_z - \nabla h_t^M$, and so we see that the transformation to this gauge involves a diffeomorphism $\psi^N_t \in \text{Diff}(M_0)$ with drive velocity ${\bf w}^N_t = -\nabla h_t^M$. This describes the motion of a surface in the quotient space $\text{Memb}$, where the velocity is only ever in the normal direction. This is illustrated in the rightmost panel of Fig.~\ref{fig3}.

\subsection{Frame Fields}
To express vector and tensor quantities on the deforming surface $M_t$ we choose a frame field ${\bf e}_1, {\bf e}_2, {\bf n}$ spanning the bundle $E_t$ along $M_t$, where ${\bf n}$ is the unit normal and ${\bf e}_j$ is a tangential frame field on $M_t$. Given that our deformable surface is moving in time, our frame will also vary in time. The most natural way to define the tangential frame is to pick a fixed set of coordinates $x_1, x_2$ on the base space $B$ and push the coordinate directions ${\bf e}_{x_j}$, $j=1,2$, forwards along the embedding $r_t$ to give a frame for the tangent space to $M_t$,
\begin{equation}
    {\bf e}_j := r_{t*} {\bf e}_{x_j}. 
\end{equation}
Conversely, any tangential frame field specified on $M_t$ can be pulled back to give a frame field on $B$. The gauge freedom is then associated with the freedom to pass to a different frame field on $B$ (or equivalently $M_t$). 

A frame field obtained by pushing forward a constant coordinate basis is what~\cite{torres-sanchez_modelling_2019} refer to as a `Lagrangian parameterisation'. This is because the frame freely moves along with the flow field, and therefore it is conceptually `Lagrangian', although we emphasise that it has nothing to do with the Eulerian--Lagrangian split in the formulation of hydrodynamics we outlined in Section \ref{sec:eulerian_vs_lagrangian}. The tangent part of this frame field is advected along with the velocity field of the surface according to 
\begin{equation} \label{eq:time_derivative}
    \partial_t {\bf e}_j = -L_{\bf v} {\bf e}_j. 
\end{equation}
We are of course free to choose a different time-dependent frame $\bar{\bf e}_j(t)$ on $B$ (which may or may not be associated with time-dependent coordinates $x_j(t)$) and push it forward to give a frame ${\bf e}_j := r_{t*} \bar{\bf e}_j$ on the surface. The choice is completely arbitrary; this is what~\cite{torres-sanchez_modelling_2019} \&~\cite{sahu_arbitrary_2020} refer to as an `arbitrary mixed Eulerian--Lagrangian parameterisation'. The time evolution is then 
\begin{equation} \label{eq:time_derivative2}
    \partial_t {\bf e}_j = -L_{\bf v} {\bf e}_j + r_{t*} \partial_t \bar{\bf e}_j.
\end{equation}
There is a special case where the frame on $B$ is chosen so that $r_{t*}\partial_t \bar{\bf e}_j = -\nabla_i v^j {\bf e}_j$; that is, the frame field is not advected by the tangential part of the flow. This is the standard choice for hydrodynamics in a fixed space. For a deforming surface, this gives rise to the concept of a frame that is advected only by the normal part of the flow, which conceptually corresponds to an observer that is `standing still' on the deforming surface. \cite{torres-sanchez_modelling_2019} refer to this as an `Eulerian parameterisation'. 

In the language of group actions, this freedom to choose the tangential part of the frame field is associated with the action of $\text{Diff}(B)$ on the frame bundle $\text{F}B \to B$, where the diffeomorphism $\psi \in \text{Diff}(B)$ acts on a frame by pushforward. Because our surface is evolving in time we consider time-dependent paths of frame fields defined by the pushforward along a time-dependent diffomorphism $\psi_t$. The gauge transformations associated to parameterisations of the surface in the `Monge gauge' and `normal gauge' that we introduced in the previous subsection act on $\text{F}B$ to produce a frame field that moves only in the $z$ direction and only in the normal direction respectively; the latter is of course the `Eulerian parameterisation' of~\cite{torres-sanchez_modelling_2019}. An arbitrary time-dependent change of gauge then gives a `mixed Eulerian--Lagrangian parameterisation'~\citep{torres-sanchez_modelling_2019, sahu_arbitrary_2020}, of which the Monge gauge is a particular example.

\section{Objective Rates}
\label{sec:objective_rates}
We now consider questions of objectivity and objective rates in fluid dynamics. Physical quantities are not generally required to be invariant under general diffeomorphisms, as a change to a noninertial reference frame will introduce `fictitious forces' as we described in the previous section---a familiar example is the Coriolis force experienced by a rotating observer. However, proper formulation of physical laws requires them to be invariant under an appropriate subgroup of diffeomorphisms, which gives rise to the concept of objectivity. Which group of transformations is required by objectivity is determined by the physics of the object under consideration.

For hydrodynamics in a fixed space $S$, a rate $D_{{\bf v}_t}$ is a way of taking the derivative of material quantities along the Eulerian flow field ${\bf v}_t$. To be a rate, the map $D$ must satisfy the Liebniz rule: for any time-dependent tensor field ${\bm \sigma}_t$ and function $f_t$ of time,
\begin{equation} \label{eq:leibniz}
    D(f_t{\bm \sigma}_t) = (Df_t){\bm \sigma}_t + f_t D {\bm \sigma}_t.
\end{equation}
This condition implies that all rates will act the same on scalar functions. Intuitively, the distinction between the ordinary time derivative $\partial_t$ and a rate $D$ is that the former only captures the way in which a material quantity itself changes in time, while the latter captures this as well as accounting for its advection along the flow.  

There is enormous freedom in defining rates. There are three notions of differentiation that make sense on a manifold---the Lie derivative, the exterior derivative, and covariant derivatives---and all of them may appear in a rate in various forms. The first two depend only on the manifold topology and are essentially unique, while there are many possible choices for the covariant derivative---requiring metric compatibility and torsion-freeness fixes the Levi-Civita connection $\nabla$ of the metric, but \textit{a priori} neither of these conditions is required to define a rate. Typical rates appearing in the literature include the Oldroyd rate~\citep{oldroyd_formulation_1950},
\begin{equation} \label{eq:oldroyd}
    D^\text{O}{\bm \sigma}_t = \partial_t{\bm \sigma}_t + L_{{\bf v}_t}{\bm \sigma}_t,
\end{equation}
the material derivative,
\begin{equation} \label{eq:material}
    D^\text{M}{\bm \sigma}_t = \partial_t{\bm \sigma}_t + \nabla_{{\bf v}_t}{\bm \sigma}_t,
\end{equation}
and the Jaumann rate~\citep{jaumann_geschlossenes_1911, prager_elementary_1961, masur_definition_1961}, also called the corotational derivative,
\begin{equation} \label{eq:jaumann}
    D^\text{J}{\bm \sigma}_t = \partial_t{\bm \sigma}_t + \frac{1}{2}\left( L_{{\bf v}_t}{\bm \sigma}_t + \left(L_{{\bf v}_t}{\bm \sigma}_t^\flat \right)^\sharp \right).
\end{equation}
Invariance of a rate under the action of a group of diffeomorphisms $G$ of the space $S$ in which the fluid moves is expressed in the following condition: the rate $D$ of any tensor ${\bf \sigma}_t$ has to satisfy the equation
\begin{equation} \label{eq:invariance}
    \phi^*  D_{\phi_* {\bf v}_t} \phi_* {\bm \sigma}_t =  D_{{\bf v}_t} {\bm \sigma}_t,
\end{equation}
for every element $\phi \in G$ of the group. Classically, objectivity of a rate is formulated as invariance under the group of Galilean transformations $\text{Gal}$, which consists of all time-independent isometries of $S$ along with time-dependent isometries that have a constant velocity, Galilean boosts.

The Lie derivative satisfies \eqref{eq:invariance} for any diffeomorphism $\phi$, which means that the Oldroyd rate \eqref{eq:oldroyd} is `general covariant'. The material derivative \eqref{eq:material} satisfies \eqref{eq:invariance} for an isometry $\phi$ provided the covariant derivative is metric compatible, i.e. $\nabla$ is the Levi-Civita connection. The Jaumann rate, being expressed in terms of Lie derivatives, also satisfies \eqref{eq:invariance} for all isometries, but unlike the Oldroyd rate it does not satisfy \eqref{eq:invariance} for a general diffeomorphism.

Now we return to the motion of a deformable surface. In this setting the rate is not a derivative on the ambient space $\mathbb{R}^3$, rather, we take the rate of quantities defined along a path $r_t$ of embeddings. The rates depend on both the path and the velocity fields associated to the tangent direction ${\bf U}_t = \partial_t r_t$ to the path. In fluid dynamics on a fixed surface possible objective rates are induced by covariant derivatives on the diffeomorphism group; mathematically this is the correct description from the perspective of the geometric theory of fluid dynamics~\citep{arnold_vladimir_2014, arnold_topological_2021}. The natural extension of this theory to the motion of deformable surfaces is to consider the configuration space $\text{Met}(B)$ of Riemannian metrics on $B$~\citep{rougee_intrinsic_2006, fiala_geometrical_2011}. This is equivalent to considering dynamics in the quotient space $\text{Memb}$, with the motion now viewed as the path $g_t = r^*_t e$ of metrics on $B$ induced by the embeddings (note that distinct paths in $\text{Emb}$ that descend to the same path in $\text{Memb}$ induce the same metrics). The tangent and cotangent spaces of $\text{Met}(B)$ play host to the strain rate and stress in this theory---as the corresponding tangent and cotangent spaces to $\text{Diff}(B)$ do for fluid dynamics in a fixed space~\citep{arnold_topological_2021}---and covariant derivatives on $\text{Met}(B)$ correspond exactly to the possible objective rates~\citep{kolev_objective_2021}. This correspondence implies that there are an enormous number of objective rates which are all equally valid from a purely mathematical point of view; physical intuition is required to choose the correct one. 

We will not adopt this perspective here. Instead, we describe the dynamics in terms of the motion in the ambient space, and objectivity in terms of invariant under a subgroup $G \subset \text{Diff}(\mathbb{R}^3)$ of the diffeomorphism group in the ambient space. Technically we should consider an appropriate subgroup of the group of diffeomorphisms of space-time, but this is equivalent to considering time-dependent paths of diffeomorphisms $\phi_t \in \text{Diff}(\mathbb{R}^3)$, and we will not labour the distinction. Under a change of coordinates corresponding to such a time-dependent paths of diffeomorphisms $\phi_t$ the velocity field ${\bf v}_t$ changes to $\phi_{t*}{\bf v}_t + {\bf w}_t$, where ${\bf w}_t = (\partial_t \phi_t) \circ \phi^{-1}_t$ is the drive velocity. The condition of invariance of a rate $D$ under this change of coordinates is
\begin{equation} \label{eq:invariance_time}
    \phi_t^*  D_{\phi_{t*} {\bf v}_t + {\bf w}_t} \phi_{t*} {\bm \sigma}_t =  D_{{\bf v}_t} {\bm \sigma}_t.
\end{equation}
Satisfying this condition requires the rate to account for, and in some sense counteract, the drive velocity of the diffeomorphism. 

To understand how this constrains the form of the rate, we explain how to construct rates starting from the most simple candidate, the partial derivative $\partial_t$ with respect to time, and describe the groups of diffeomorphisms that they are invariant under. Under the action of a diffeomorphism $\phi_t$ the rate of a tensor $\sigma_t$ changes to
\begin{equation}
    \begin{aligned}
        \partial_t \phi_{t*} {\bm \sigma}_t &= \phi_{t*}\partial_t {\bm \sigma}_t - L_{{\bf w}_t} \phi_{t*} {\bm \sigma}_t, \\
            &= \phi_{t*}\left( \partial_t {\bm \sigma}_t - L_{\phi_{t}^* {\bf w}_t} {\bm \sigma}_t \right).
    \end{aligned}
\end{equation}
In the second line we have used the facts that $\phi^*\phi_* = \phi_* \phi^*$ is the identity that that $\phi_* L_{\bf u} \sigma =  L_{\phi_* {\bf u}} \phi_* {\bm \sigma}$ for any diffeomorphism $\phi$, vector field ${\bf u}$, and tensor ${\bm \sigma}$. This calculation shows that the partial derivative is only invariant under time-independent diffeomorphisms. This calculation motivates the definition of the Oldroyd rate (\ref{eq:oldroyd}), as the addition of a Lie derivative causes the extra terms that involve the drive velocity to cancel,
\begin{equation}
    \begin{aligned}
        \partial_t (\phi_{t*} {\bm \sigma_t}) + L_{\phi_{t*}{\bf v}_t + {\bf w}_t} (\phi_{t*} {\bm \sigma_t}) &= \phi_{t*}\left( \partial_t {\bm \sigma_t} - L_{\phi_{t}^* {\bf w}_t}  {\bm \sigma_t} + L_{{\bf v}_t} {\bm \sigma_t} + L_{\phi_t^* {\bf w}_t} {\bm \sigma_t} \right),  \\
        &= \phi_{t*}\left( \partial_t {\bm \sigma_t} + L_{{\bf v}_t} {\bm \sigma_t}  \right),
    \end{aligned}
\end{equation}
and thus the Oldroyd rate is invariant under all time-dependent diffeomorphisms. By contrast, the material derivative (\ref{eq:material}) is not invariant under general diffeomorphisms, or even general time-dependent isometries. Indeed, we compute that 
\begin{equation} \label{eq:pushforward_material}
    \partial_t (\phi_{t*} {\bm \sigma}_t) + \nabla_{\phi_{t*}{\bf v}_t + {\bf w}_t} (\phi_{t*} {\bm \sigma}_t) = \phi_{t*}\left( \partial_t {\bm \sigma}_t - L_{\phi_{t}^* {\bf w}_t}  {\bm \sigma}_t + \nabla_{{\bf v}_t} {\bm \sigma}_t + \nabla_{\phi_t^* {\bf w}} {\bm \sigma}_t \right). 
\end{equation}
When $\phi_t $ is a translation with constant velocity, then covariant derivatives equal Lie derivatives $\nabla_{\phi_t^* {\bf w}} {\bm \sigma}_t = L_{\phi_t^* {\bf w}} {\bm \sigma}_t$. Using this relation in \eqref{eq:pushforward_material} causes the extra terms to cancel and illustrates that the material derivative is objective when we restrict our transformations to constant rotations and time-dependent translations with constant velocity. The latter are Galilean boosts, and hence this calculation shows that the material derivative is invariant under exactly the Galilean group $\text{Gal}$.

We argue that it is a physical necessity that rates satisfy an additional condition beyond objectivity: the Euclidean metric $e$ in the ambient space should not be advected by the flow,
\begin{equation} \label{eq:noadvection}
    D_{{\bf v}_t} e = 0. 
\end{equation}
This asserts that the motion of the fluid does not change the fundamental structure of space time. Aside from being an obvious physical necessity, this condition is important to avoid the appearance of changes in the induced metric on the surface that are not related to a real deformation. Consider the following simple example. The induced metric (first fundamental form) $g$ on the surface is defined by taking the Euclidean metric and projecting it down onto the surface. The rate of the surface metric is therefore
\begin{equation} \label{eq:rate_of_metric}
    D g = D e - {\bf n}^\flat \otimes D {\bf n}^\flat - D {\bf n}^\flat \otimes {\bf n}^\flat, 
\end{equation}
where ${\bf n}$ is the surface normal. Suppose we have a flat surface $z=0$ undergoing a purely tangential flow. The surface does not change, and hence the metric induced on the surface is constant in both space and time, as is the normal---thus, there should be no advection with the flow at all. This would also be the case for a flow normal to the surface but spatially constant, meaning the surface simply moves without deforming. If the rate advects the Euclidean metric then \eqref{eq:rate_of_metric} implies an advection of either the metric or the normal. Plainly this should not be the case for the examples just given where there is no deformation. 

Lie derivatives are general covariant but do not preserve the metric except in the special case when the velocity field is a Killing field. Typically this will not be the case, and hence the Oldroyd rate is out if we require this additional condition. The partial derivative clearly does preserve the ambient metric, but lacks the appropriate objectivity properties. The material derivative obviously preserves the metric---provided it involves a metric-compatible connection---but is not invariant under general isometries. 

The requirement of preserving the metric as well as being invariant under all time-dependent isometries motivates the addition of a corotational part to the Oldroyd rate, which leads us to the Jaumann rate (\ref{eq:jaumann}). 
The corotational terms must themselves be invariant under a time-dependent isometry. Eq.~\eqref{eq:pushforward_material} illustrates that these terms cannot involve covariant derivatives, only combinations of Lie derivatives~\citep{marsden_mathematical_1994}. The simplest possible expression with this property is then 
\begin{equation}
\frac{1}{2}\left( \left(L_{{\bf v}_t}{\bm \sigma}_t^\flat \right)^\sharp - L_{{\bf v}_t}{\bm \sigma}_t\right).
\end{equation}
Adding this term to the Oldroyd rate indeed yields the Jaumann rate ~\eqref{eq:jaumann}. Because these terms involve the metric connection (through the raising and lowering operators) the Jaumann rate is not invariant under a general diffeomorphism, but it is invariant under all time-dependent paths of isometries. Furthermore, the Jaumann rate satisfies the condition~\eqref{eq:noadvection} and does not advect the ambient metric. The result of these calculations for the different rates is summarised in Table \ref{tab:invariant}.

\begin{table}
    \begin{center}
        \begin{tabular}{lcc}
            Rate & Objective with respect to & Preserves ambient metric?\\
            \hline
            Partial derivative \, \, \, & time-independent $\phi \in \text{Diff}(\mathbb{R}^3)$ & Yes\\ 
            Oldroyd & time-dependent $\phi_t \in \text{Diff}(\mathbb{R}^3)$ & No\\ 
            Material & Galilean transforms $\phi_t \in \text{Gal}$ & Yes\\ 
            Jaumann & time-dependent $\phi_t \in \text{E}(3)$ & Yes\\ 
        \end{tabular}
        \caption{For the four rates considered in the main text we show the largest group of diffeomorphisms under which the given rate is invariant. We also indicate whether the rate preserves the Euclidean metric in the ambient space.}
        \label{tab:invariant}
    \end{center}
\end{table}

It follows that the plethora of rates in the literature can be reduced to two, the material and the Jaumann. Which of these two rates is the correct one? The essential difference between them is that the material derivative `feels' constant speed rotations while the Jaumann is not. The correct choice of rate for a given material quantity then boils down to whether that quantity should be sensitive to angular velocities on the deforming surface. In the gauge transformation picture outlined in the previous section this can be interpreted as a question about which of the spaces $\text{Emb}, \text{Memb}$, and $\text{Iso}$ the quantity lives in. 

A quantity (by which we mean physical fields as well as differential operators like rates) defined on $\text{Emb}$ can be `pushed down' to one of the quotient spaces $\text{Memb}$ and $\text{Iso}$ by removing the part of it that changes under the action of the appropriate gauge group. A quantity can be defined on $\text{Iso}$ if it only depends on the first and second fundamental forms of the surface and is insensitive to a spatially-constant but time-dependent angular momentum, which has no meaning when we quotient out by the action of the isometry group. Concretely, if ${\bf v}$ is the velocity field of our surface, we can decompose it as ${\bf v} = {\bf v}^\text{ess} + {\bf w}^\text{iso}$, where ${\bf w}^\text{iso}$ is the drive velocity of some time-dependent isometry and ${\bf v}^\text{ess}$ is the `essential' part. The Jaumann rate depends only on this essential part and ignores the isometry part---this is equivalent to the rate descending to a well-defined derivative on the quotient space $\text{Iso}$. The key example of a quantity that should not feel the effect of a time-dependent isometry is a nematic director or Q-tensor in the deforming surface. Such a nematic director will not feel a translation nor a continuous rotation of the surface about an axis, and thus its rate should be invariant under any time-dependent isometry. The Jaumann rate is the appropriate choice for any physical quantity which has this property, and indeed corotational derivatives are used to formulate the equations of nematodynamics in a flat space~\citep{degennes_physics_2013,stewart_static_2019}. 

A physical quantity which does not have this property is the surface velocity. For example, a spherical surface experiencing a constant speed rotation about an axis (a time-dependent isometry, but not an element of the Galilean group) will experience a centripetal force inwards that will act to deform it. Velocities do care about ambient space motions, and hence we should choose a rate that involves invariance under the Galilean group but not the full group of time-dependent isometries. The correct rate for a velocity field is therefore the material rate, not the Jaumann. The appropriate rate for a general physical quantity in morphodynamics can be deduced analogously by determining its behaviour under rotations with constant speed.

\section{Computations of the Material and Jaumann Rate}
\label{sec:formulas}

We now give expressions for the rates of various quantities that are important for membrane dynamics and morphology. Throughout this section we denote the Jaumann rate along the Eulerian velocity ${\bf v}_t$ by $D^\text{J}_t$ and the material derivative by $D_t^\text{M}$. As described in the previous section, the material derivative is the appropriate rate for velocities, while the Jaumann is the appropriate rate for quantities that are insensitive to angular momentum, such as a nematic Q-tensor or polarisation field. We also give two expressions, one in a coordinate basis and and a second where the quantities are computed relative to an orthonormal frame.  

In Appendix~\ref{appA} we give a clear derivation of these formulas using Cartan's method of moving frames in the ambient space. In Appendix~\ref{appB} we give an alternative derivation using the surface covariant derivative. Both approaches are equivalent and give the same result. 

\subsection{Rates in a Coordinate Frame}
We begin with the formulae in a coordinate frame. Let $x_j$, $j=1,2$, denote coordinates on $B$. We push the associated coordinate basis ${\bf e}_{x_j}$ on $B$ forward along the embeddings $r_t$ to obtain a coordinate basis on $M_t$, 
\begin{equation}
    {\bf e}_j := r_{t*} {\bf e}_{x_j}. 
\end{equation}
These will not in general be orthogonal or normalised. We follow~\cite{al-izzi_morphodynamics_2023} and choose a time-varying coordinate frame on $B$ that is not advected by the left-Eulerian velocity ${\bf V}_t$. We make this choice to ensure that our formulas are consistent with those derived for ordinary hydrodynamics in a flat space, and also because this is the most practical choice for computations.

Write ${\bf n}$ for the surface normal. The velocity field is ${\bf v} = v^j {\bf e}_j + v^n {\bf n}$. Let ${\bf u} = u^j {\bf e}_j + u^n {\bf n}$ be some other vector field. Its material derivative is
\begin{equation} \label{eq:covariant}
    \begin{aligned}
        D_t^\text{M} {\bf u} &= \left( \partial_t u^j + v^i \nabla_i u^j + u^iv^k \omega_{ki}^j -u^iv^nb_i^j -v^iu^n b_{i}^j - u^n\nabla^j v^n \right){\bf e}_j \\
        & \ \ \ \ + \left(\partial_t u^n + v^i\nabla_i u^n + v^iu^jb_{ij} + u^i\nabla_i v^n \right){\bf n}.
    \end{aligned}
\end{equation}
This leads to a formula for the acceleration, 
\begin{equation} \label{eq:coordinate_acceleration}
    \begin{aligned}
        D_t^\text{M} {\bf v} &= \left( \partial_t v^j + v^i \nabla_i v^j + v^iv^k \omega_{ki}^j -v^iv^nb_i^j -v^iv^n b_{i}^j - v^n\nabla^j v^n \right){\bf e}_j \\
        & \ \ \ \ + \left(\partial_t v^n + v^i\nabla_i v^n + v^iv^jb_{ij} + v^i\nabla_i v^n \right){\bf n}.
    \end{aligned}
\end{equation}
The details of these calculations are shown in Appendices \ref{appA} and \ref{appB}. In these and all other formulas in this section, $\nabla$ denotes the Euclidean metric connection in the ambient space, not the connection on the surface, and $\omega_{ij}^k$ are components of the connection 1-form as described in Appendix \ref{appA}. A discussion of the relationship between the ambient connection and the surface is given in Appendix \ref{appB}. This formula for the acceleration is identical to the one derived in \cite{waxman_dynamics_1984} \& \cite{scriven_dynamics_1960} by less formal means.

\begin{figure}
 \centering
 \includegraphics[width=0.3\linewidth]{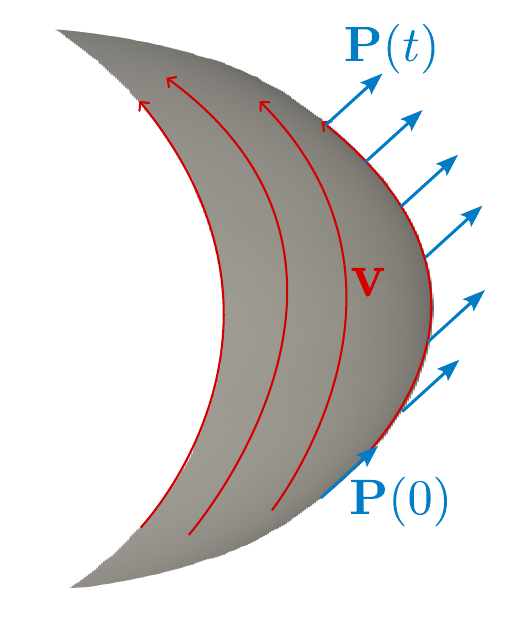}
 \caption{Lie dragging a tangent vector field ${\bf P}$ along a flow ${\bf v}$ on the surface results in a new vector field which has a component out of the surface--we illustrate ${\bf P}$ at a single point being dragged along an integral curve of ${\bf v}$. In order to keep the field ${\bf P}$ in the surface, a counteracting force must push back against this effect.}
 \label{fig4}
\end{figure}

We also compute the Jaumann rate of ${\bf u}$,
\begin{equation} \label{eq:jaumann}
    \begin{aligned}
        D_t^\text{J} {\bf u} &= \left(\partial_t u^j + v^i\nabla_i u^j -u^iv^nb_i^j - \frac{1}{2}v^iu^nb_i^j -\frac{1}{2}u^n\nabla_jv^n - \frac{u^i}{2}(\nabla_i v^j - \nabla^jv_i) \right){\bf e}_j \\
        & \ \ \ \ + \left(\partial_t u^n + \frac{1}{2}u^i\nabla_i v^n + v^i \nabla_i u^n + \frac{1}{2}v^iu^jb_{ij}\right){\bf n}.
    \end{aligned}
\end{equation}
By setting the normal component of ${\bf u}$ to zero in \eqref{eq:jaumann}, we obtain the Jaumann rate of a nematic polarisation field ${\bf P} = P^1 {\bf e}_1 + P^2 {\bf e}_2$ tangent to the surface
\begin{equation} \label{eq:jaumann_director}
    \begin{aligned}
        D_t^\text{J} {\bf P} &= \left(\partial_t P^j + v^i\nabla_i P^j -P^iv^nb_i^j - \frac{P^i}{2}(\nabla_i v^j - \nabla^jv_i) \right){\bf e}_j \\
        & \ \ \ \ + \frac{1}{2}\left(P^i\nabla_i v^n + v^iP^jb_{ij}\right){\bf n}.
    \end{aligned}
\end{equation}
Note this formula has a normal component, which may seem surprising. However, we can see this as the predictable result of Lie-dragging a vector field along a curved surface. Consider a flow that is tangent to the surface, and drag a vector ${\bf P}(0)$ at a given point along that flow as in Fig.~\ref{fig4} to get a new vector ${\bf P}(t)$. We see that this results in the vector lifting off the surface.

Keeping a vector field confined to the tangent plane of a surface therefore requires additional forces. These might arise from an embedding fluid, or from some other, unspecified interfacial physics, but a mathematical understanding of the precise nature of such forces typically requires a consideration of how finite thickness surfaces formally limit to manifolds, e.g.,~\citep{lomholt_descriptions_2006, lomholt_fluctuation_2006, lomholt_mechanics_2006}. Either way, we can imagine a counteracting force whose strength is sufficient to ensure that the nematic director being tangential is a hard constraint. This can be implemented using a Lagrange multiplier, or simply by projecting \eqref{eq:jaumann_director} into the surface, which results in the formula derived elsewhere~\citep{salbreux_theory_2022, al-izzi_morphodynamics_2023, degennes_physics_2013,stewart_static_2019}.

\subsection{Rates in an Orthonormal Frame}
An alternative approach to the use of a coordinate frame is to work with an orthonormal basis ${\bf e}_1, {\bf e}_2$ for the deforming surface. This approach may be a convenient choice when dealing with a nematic polarisation field on the surface, as we can take this to be the basis vector ${\bf e}_1$. 

For a general basis, the time derivatives are 
\begin{equation}
    \begin{aligned}
        \partial_t {\bf e}_j &= R_t J{\bf e}_j + \nabla_j v^n {\bf n}, \\
        \partial_t {\bf n} &= -\nabla^j v^n {\bf e}_j.  
    \end{aligned}
\end{equation}
Here, we have introduced the map $J = {\bf n} \times$ which affects a 90 degree rotation about the unit normal, $J{\bf e}_1 = {\bf e}_2$ and $J{\bf e}_2 = -{\bf e}_1$. The function $R_t$ is a time-dependent function which characterises the in-plane rotation of the basis. The choice of this function is a gauge freedom, and it may be set to zero by a particular choice of the ON frame. 

The equation for the material derivative is 
\begin{equation} \label{eq:covariant_ON}
    \begin{aligned}
    D^\text{M}_t {\bf u} &= \left(\partial_t u^k + v^i\nabla_i u^k + v^iu^j\omega_{ij}^k - u^n\nabla^k v^n - v^iu^nb_i^k \right){\bf e}_k + u^kR_t J {\bf e}_k \\
    & \ \ \ \ \ + \left(\partial_t u^n + v^i \nabla_i u^n + u^j\nabla_j v^n + v^i u^jb_{ij}\right) {\bf n}.
    \end{aligned}
\end{equation}
The Jaumann rate of a vector field is,
\begin{equation} \label{eq:jaumann_ON}
    \begin{aligned}
        D^\text{J}_t{\bf u} &= u^jR_t J{\bf e}_j + \bigg(\partial_t u^k + v^i \nabla_i u^k + \frac{u^j}{2}\left(\nabla^k v_j - \nabla_j v^k\right) + \frac{v^iu^j}{2}(2\omega_{ij}^k + \omega_{jk}^i -\omega_{kj}^i) \\
         & \ \ \ \ \ -\frac{u^n}{2}(\nabla^k v^n + v^jb_j^k)\bigg){\bf e}_k  + \left(\partial_t u^n + v^i \nabla_i u^n +\frac{1}{2}v^iu^j b_{ij} + \frac{1}{2}u^j\nabla_j v^n\right){\bf n}.
    \end{aligned}
\end{equation}
Again, these equations use the ambient space connection $\nabla$ and not the surface connection. Written out in this choice of frame, the rate now involve the connection coefficients $\omega_{ij}^k$, which are not be symmetric in the lower two indices as they are for a coordinate frame. If we do choose the component ${\bf e}_1$ of the frame to be a nematic polarisation direction, then these coefficients have the physical interpretation as the surface splay and bend distortions of the polarisation field. Write $\kappa$ for the (in-plane) bend magnitude and $s$ for the splay of the polarisation field. Then the connection coefficients are~\citep{pollard_intrinsic_2021, da_silva_moving_2021}, 
\begin{equation}
    \omega_{11}^2 = -\omega_{12}^1 = \kappa, \ \ \ \ \ \ \ \omega_{21}^2 = -\omega_{22}^1  = s.
\end{equation}
We can use these to express~\eqref{eq:jaumann_ON} in terms of the nematic distortion modes.

\section{Discussion}
In this paper, we have brought some mathematical clarity to the formulation of morphodynamics. We have described the Eulerian and Lagrangian perspectives on the motion of a fluid deformable surface and shown how these map on to equivalent formulations for the motion of a fluid in a fixed space. Using a symmetry group analysis, we have shown that morphodynamics has three inherent (and independent) gauge freedoms,  which are associated with isometries of the ambient space, the parametrisation of the surface, and the choice of the frame field.

Ultimately, once we have disambiguated between these notions, we argue that objective rates in morphodynamics should be treated precisely as with their fixed space counterparts: as invariant under time-dependent isometries of the ambient (full) space. As shown elsewhere, however, this yields an infinite number of possibilities \citep{kolev_objective_2021}. We therefore require that these rates should also conserve the ambient space metric, as they would in a fixed space. Under this additional constraint, we show that the material and Jaumann rates are the only possible choices.

The material derivative leads to rates that are invariant under the action of the Galilean group--- fixed isometries and constant velocity translations, or `boosts'. Since Noether's theorem identifies the Galilean group with conservation of momentum, this therefore applies to the velocity field; the resultant rate (i.e., acceleration) is not invariant under non-inertial gauges, which induce fictitious forces. The Jaumann derivative, by contrast, is invariant under {\it all} time-dependent isometries. It is the correct rate for hydrodynamic variables whose long wavelength behaviour arises not from conservation laws, but due to symmetry-breaking, and the related presence of Goldstone modes~\citep{hohenberg_theory_1977}: a pertinent example is the $Q$-tensor that captures nematic degrees-of-freedom.

Taken together, this protocol essentially ensures that the largest group of time-dependent isometries that the whole system is invariant under is just the Galilean group, as required by the informal understanding of objectivity. We provide several examples of how to calculate such rates, which includes two popular choices of frame field.

One point of contention that this raises is with the use of the Oldroyd rate in constitutive relations of certain non-Newtonian fluids~\citep{oldroyd_formulation_1950,edwards_oldroyds_2023}. It is not clear to us how this practice fits into the framework that we have outlined.

More generally, we remark that many of the ideas that we present borrow from, or are motivated by, Arnold's theory of hydrodynamics~\citep{arnold_topological_2021}, which has been greatly influential and has led to a deeper understanding of fluid dynamics, especially the role played by topology and geometry. To take the analogy further would involve a formulation of morphodynamics which focuses on the Riemannian metric and second fundamental form as the central dynamical objects, as opposed to the embedding. Such an approach has been discussed previously by~\cite{morris_active_2019}, where both objects were treated as symmetry-broken variables, but to our knowledge the idea not been developed since. Aside from offering a fresh perspective, computer simulations in this formulation of morphodynamics seemingly sidestep computational issues which require the adoption of a normal gauge or a Monge gauge. Either way, we argue that geometrical tools, such as the pushforward and pullback, gauge symmetry analysis, and the language of differential geometry and/or exterior calculus can help to highlight and clarify the role played by geometry in the morphodynamics of fluid deformable surfaces, and we welcome further work in the area.

\appendix

\section{Calculation of Rates in the Cartan Formalism}\label{appA}
We calculate the material and Jaumann rates using Cartan's moving frame approach to differential geometry~\citep{cartan_systemes_1945,frankel_geometry_2011}. In what follows, Greek indices $\alpha, \beta, \gamma$ run from from 1 to 3, while Latin indices $i,j,k$ run from 1 to 2. Let ${\bf e}_\alpha$ be any frame along the surface, with ${\bf e}_3 =  {\bf n}$ the normal direction. Write $e^\alpha$ for the dual 1-form. Define the connection 1-forms associated to the Euclidean connection $\nabla$ in the ambient space by 
\begin{equation}
    \begin{aligned}
        \nabla {\bf e}_\alpha = \omega^\beta_\alpha {\bf e}_\beta, \\
        de^\alpha = -\omega_\beta^\alpha \wedge e^\beta
    \end{aligned}
\end{equation}
with components,
\begin{equation}
    \omega_\beta^\gamma({\bf e}_\alpha) = \omega_{\alpha\beta}^\gamma.
\end{equation}
In our problem it doesn't make sense to take derivatives along the normal ${\bf n}$, so the surface geometry is encoded by dropping all coefficients of $e^3$ from the connection form. The 1-forms $\omega_3^1, \omega_3^2$ then encode exactly the second fundamental form $b = -\nabla {\bf n}$, while $\omega_1^2$ is the connection 1-form of the induced metric on the surface:
\begin{equation}
    \begin{aligned}
      \omega_3^1 &= -b^1_{1}e^1 - b^1_{2}e^2, \\
      \omega_3^2 &= -b_{1}^2 e^1 - b_{2}^2e^2, \\
      \omega_1^2 &= \omega_{11}^2 e^1 + \omega_{21}^2e^2.
    \end{aligned}
\end{equation}
We now separate out the normal component of the frame from the tangential components. Then the connection 1-form encodes the following relationships, 
\begin{equation}
    \begin{aligned}
        \nabla_i {\bf e}_j &= \omega_{ij}^k{\bf e}_k + b_{ij}{\bf n}, \\
        \nabla_i {\bf n} &= -b_{i}^j{\bf e}_j.
    \end{aligned}
\end{equation}
We are free to choose the tangential frame however we wish. Following~\cite{al-izzi_morphodynamics_2023} we take the components ${\bf e}_1, {\bf e}_2$ of the frame to be the pushforward of some coordinates $x_1,x_2$ on the base space $B$ along the embedding $r$. Explicitly, we define coordinate vectors ${\bf e}_{x_j}$ on the base $B$, and then set
\begin{equation}
    {\bf e}_j := r_{*} {\bf e}_{x_j}. 
\end{equation}
Using the rules for the time derivative of a pushforward, see Appendix \ref{appC}, along with the fact that coordinates $x_1, x_2$ do not depend on time, we conclude that 
\begin{equation} \label{eq:time_derivative}
    \partial_t {\bf e}_j = -L_{\bf v} {\bf e}_j,
\end{equation}
where ${\bf v} = \partial_t r \circ r^{-1}$ is the velocity field. Moreover, since the $x_j$ are coordinates we have 
\begin{equation}
    [{\bf e}_1, {\bf e}_2] = [r_*{\bf e}_{x_1}, r_*{\bf e}_{x_2}] = r_*[{\bf e}_{x_1}, {\bf e}_{x_2}] = 0.
\end{equation}
This implies $(\omega_{12}^k - \omega_{12}^k) = 0$, and so the coefficients in the components of the connection 1-forms are all symmetric. The time derivative of the normal component can be derived from the fact that it is orthogonal to the ${\bf e}_j$, which implies that ${\bf n} \cdot \partial_t{\bf e}_j = - {\bf e}_j \cdot \partial_t {\bf n}$. 

Write the velocity as ${\bf v} = v^j {\bf e}_j + v^n {\bf n}$. To compute the rate of any vector field ${\bf u} = u^\alpha {\bf e}_\alpha$, we can use the Leibniz formula,
\begin{equation} \label{eq:leibniz}
    D_t {\bf u} = \left(\partial_t u^\alpha + v^\beta\nabla_\beta u^\alpha\right){\bf e}_\alpha + u^\alpha D_t{\bf e}_\alpha.
\end{equation}
In order to compute the material and Jaumann rates, we need only compute their action on the basis elements. Firstly, let us compute the time derivatives of the frame. By \eqref{eq:time_derivative}, we obtain this from the Lie derivatives, 
\begin{equation} \label{eq:lie_derivatives_frame}
    \begin{aligned}
        L_{\bf v}{\bf e}_j &= \nabla_{\bf v} {\bf e}_j - \nabla_{{\bf e}_j} {\bf v}, \\
                           &= v^i\left(\omega_{ij}^k - \omega_{ji}^k \right){\bf e}_k - \nabla_j v^i{\bf e}_i - v^n \nabla_j {\bf n} - \nabla_j v^n{\bf n}, \\
                           &= \left(v^n b_j^i - \nabla_j v^i \right) {\bf e}_i - \nabla_j v^n {\bf n} \\ 
        L_{\bf v} {\bf n}  &= \nabla_{\bf v} {\bf n}, \\
                           &= -v^jb_j^i {\bf e}_i, \\
    \end{aligned}
\end{equation}
Here we have used the fact that ${\bf e}_i$ is a coordinate basis, so that $\omega_{ij}^k$ is symmetric. These yield the time derivatives for a constant coordinate frame on the base space. In practice, it is preferable to choose a time-dependent coordinate basis for $B$, corresponding to performing a time-dependent gauge transformation. The formula for the change of the frame is then
\begin{equation} \label{eq:time_derivative2}
    \partial_t {\bf e}_j = -L_{\bf v} {\bf e}_j + r_* \partial_t {\bf e}_{x_j}.
\end{equation}
We choose the time-dependent change of gauge so that $r_* \partial_t {\bf e}_{x_j}$ cancels out the term $\nabla_j v^i {\bf e}_i$ that would appear in the Lie derivative of ${\bf v}$; informally, this involves moving to a coordinate frame that is not advected by the tangential part of the flow. This particular choice then results in the time derivatives obtained by~\cite{al-izzi_morphodynamics_2023} \& \cite{salbreux_theory_2022},
\begin{equation} \label{eq:time_derivatives}
    \begin{aligned}
        \partial_t {\bf e}_i &= - v^n b_i^j{\bf e}_j + \nabla_i v^n{\bf n} , \\
        \partial_t {\bf n} &= - \nabla^j v^n{\bf e}_j. 
    \end{aligned}
\end{equation}
We continue to work in this natural choice of time-dependent coordinate system in what follows. 

Now we compute our rates. We begin with the material rate. The action of this rate on the basis vectors is,
\begin{equation}
    \begin{aligned} 
    \partial_t {\bf e}_i + \nabla_{\bf v}{\bf e}_i &= \left(v^k \omega_{ki}^j -v^nb_i^j\right) {\bf e}_j + \left(v^jb_{ij} + \nabla_i v^n\right){\bf n}, \\
    \partial_t {\bf n} + \nabla_{\bf v}{\bf n} &= -\left(v^i b_{i}^j + \nabla^j v^n\right){\bf e}_j,
    \end{aligned}
\end{equation}
Inserting these terms in the Leibniz formula \eqref{eq:leibniz} for a general vector field ${\bf u}$ leads us to Eq. \eqref{eq:covariant}. 
We already have the Lie derivatives of the frame, Eq.~\ref{eq:lie_derivatives_frame}, so to obtain the Jaumann rate it remains to compute the Lie derivatives of the coframe,  
\begin{equation}  \label{eq:lie_derivatives_coframe}
    \begin{aligned}
        L_{\bf v}e^j &= \iota_{\bf v}de^j + dv^j, \\
                     &= \iota_v\left( -\omega^j_{ik} e^i \wedge e^k + b_i^j e^i \wedge {\bf n}^\flat \right) + \nabla_iv^j e^i, \\
                     &= \left(v^i b_i^j {\bf n}^\flat - v^nb_i^je^i + v^k(\omega_{ik}^j - \omega_{ki}^j)e^i \right)  + \nabla_i v^j e^i, \\
                     &= \left(\nabla_i v^j - v^nb_i^j \right)e^i + v^i b_i^j {\bf n}^\flat, \\
        L_{\bf v} {\bf n}^\flat &= \iota_{\bf v} d{\bf n}^\flat + dv^n, \\
                                &= \nabla_i v^n e^i.
    \end{aligned}
\end{equation}
We have again used the fact that $\omega_{ij}^k$ is symmetric for the vanishing of the tangential part of the connection form, and the 2-form $d{\bf n}^\flat$ vanishes because the second fundamental form is also symmetric. These then lead us to Lie derivatives of a general vector field ${\bf u}$ and its dual 1-form ${\bf u}^\flat$
\begin{equation}    
    \begin{aligned}
    L_{\bf v}{\bf u} &= u^j L_{\bf v} {\bf e}_j + v^i\nabla_i u^j{\bf e}_j + u^n L_{\bf v} {\bf n} + v^i\nabla_i u^n{\bf n}, \\
                     &= \left(v^i\nabla_i u^j - u^i\nabla_i v^j - u^nv^i b_i^j + v^nu^i b_i^j\right) {\bf e}_j + \left(v^i\nabla_i u^n - u^i\nabla_i v^n  \right) {\bf n}, \\
    \end{aligned}
\end{equation}
\begin{equation}    
    \begin{aligned}
    L_{\bf v}{\bf u}^\flat &= u_j L_{\bf v} {\bf e}^j + v^i\nabla_i u_je^j + u^nL_{\bf v} {\bf n}^\flat + v^i\nabla_i u^n {\bf n}^\flat, \\
                     &= \left( v^i \nabla_i u_j + u^n \nabla_j v^n + u_i \nabla_j v^i - v^nu_i b_j^i \right) e^j + \left(v^i\nabla_i u^n + u_iv^jb_j^i \right){\bf n}^\flat.
    \end{aligned}
\end{equation}
Putting these together leads to
\begin{equation} \label{eq:corotational}
    \begin{aligned}
        \frac{1}{2}\left(L_{\bf v} {\bf u} + \left(L_{\bf v} {\bf u}^\flat \right)^\sharp \right)     &= \frac{1}{2}\left(2v^i\nabla_i u^j + u^n \nabla^j v^n -v^iu^nb_i^j + u^i(\nabla^jv_i - \nabla_i v^j)\right) {\bf e}_j \\
        & \ \ \ \ \ +\frac{1}{2}\left(v^iu^jb_{ij} + 2v^i\nabla_i u^n - u^i\nabla_iv^n\right){\bf n}
    \end{aligned}
\end{equation}
Now we compute the time derivative of ${\bf u}$ using \eqref{eq:time_derivatives},
\begin{equation} \label{eq:partialtu}
    \partial_t {\bf u} = \left(\partial_t u^j - u^iv^n b_i^j -u^n \nabla^j v^n\right){\bf e}_j + \left(\partial_t u^n + u^i\nabla_i v^n \right){\bf n}.
\end{equation}
Combining Eq.~\eqref{eq:corotational} with Eq.~\eqref{eq:partialtu} then gives the final result, Eq.~\eqref{eq:jaumann}. We emphasise that this formula holds only when ${\bf e}_j$ is a specific time-dependent choice of coordinate basis of the base space $B$. For a time-independent coordinate basis on $B$ we pick up the additional term $u^i\nabla_i v^j {\bf e}_j$ that accounts for the advection of the coordinates. 

We now choose to work with an orthonormal (ON) basis ${\bf e}_1, {\bf e}_2$ for the tangent space to the the deforming surface, and introduce ${\bf e}_3 = {\bf n}$. Along the surface, the euclidean metric $e$ does not change, and therefore we have $\partial_t {\bf e}_\alpha \cdot {\bf e}_\beta = 0$. The evolution of the normal is constrained by the fact that $\partial_t {\bf n} = -\nabla^j v^n {\bf e}_j$, and therefore we have 
\begin{equation}
    \begin{aligned}
        \partial_t {\bf e}_1 &= R_t {\bf e}_2 + \nabla^1 v^n {\bf n}, \\
        \partial_t {\bf e}_2 &= -R_t {\bf e}_1 + \nabla^2 v^n {\bf n}, \\
        \partial_t {\bf n} &= -\nabla^1 v^n {\bf e}_1 - \nabla^2 v^n {\bf e}_2. 
    \end{aligned}
\end{equation}
Here, $R_t$ is a function on $M_t$ of both position and time. It describes an intrinsic rotation of the coordinate system which we are free to choose as we wish, including by setting it to be zero, as it is a gauge freedom. For reasons of generality we choose to leave it in our calculations. Introduce the operation $J = {\bf n} \times$. Then we may write more succinctly, 
\begin{equation}
    \begin{aligned}
        \partial_t {\bf e}_j &= R_t J{\bf e}_j + \nabla^j v^n {\bf n}, \\
        \partial_t {\bf n} &= -\nabla^j v^n {\bf e}_j.  
    \end{aligned}
\end{equation}
The material rate is then readily seen to be \eqref{eq:covariant_ON}. The Lie derivatives of ${\bf n}, {\bf n}^\flat$ in this ON frame are the same as in the coordinate frame. The Lie derivatives of ${\bf e}_j$ and $e^j$ are then, 
\begin{equation} \label{eq:lie_derivatives_ON_frame}
    \begin{aligned}
        L_{\bf v}{\bf e}_j &= \left(v^i(\omega_{ij}^k - \omega_{ji}^k) + v^n b_j^k - \nabla_j v^k \right) {\bf e}_k - \nabla_j v^n {\bf n} \\ 
        L_{\bf v}e^j &= \left(v^i(\omega_{ki}^j - \omega_{ik}^j) + \nabla_k v^j -v^nb_k^j \right)e^k + v^i b_i^j {\bf n}^\flat.
    \end{aligned}
\end{equation}
Thus, 
\begin{equation} \label{eq:ONframe_rates}
    \begin{aligned}
        D^\text{J}_t{\bf e}_j &= R_t J {\bf e}_j + \frac{1}{2}\Big(\nabla^k v_j - \nabla_j v^k +v^i(2\omega_{ij}^k + \omega_{jk}^i -\omega_{kj}^i)\Big){\bf e}_k + \frac{1}{2}\left(v^i b_{ij} + \nabla_j v^n\right) {\bf n}, \\
        D^\text{J}_t{\bf n} &= -\frac{1}{2}\left(\nabla^k v^n + v^j b_j^k\right) {\bf e}_k. \\
    \end{aligned}
\end{equation}
Using the Leibniz formula, we then derive the Jaumann rate of a general vector field in an ON frame,~\eqref{eq:jaumann_ON}.

\section{Calculation of Rates Using Christoffel Symbols}\label{appB}
Here we give a computation of the objective and Jaumann rates using more traditional differential geometry notation based on a coordinate basis. The Gauss and Weingarten equations are,
\begin{align}
    \partial_i{\bf e}_j &= \Gamma^k_{ij}{\bf e}_k + b_{ij}{\bf n}\text{,}\\
    \partial_i{\bf n} &= - b_{i}{}^{k}{\bf e}_k\text{,}
\end{align}
where $\Gamma^k_{ij}=\frac{1}{2}g^{kl}\left( \partial_i g_{lj} + \partial_j g_{li} -\partial_l g_{ij}\right)$ are the Christoffel symbols associated with the induced metric of the surface. The dynamics of the basis is given by,
\begin{align}
    \partial_t{\bf e}_j &= - v^n b_{i}{}^{j}{\bf e}_j + \bar\nabla_iv^n {\bf n}\text{,}\\
    \partial_t{\bf n} &= -\bar\nabla_i v^n {\bf e}_i\text{,}
\end{align}
where $\bar\nabla$ is the covariant derivative on the surface (not the ambient connection $\nabla$). Note that we have made a choice of frame here where our coordinates move only with the normal velocity, such that for a fixed flat manifold we will recover the standard formulae of fluid mechanics.

In addition we note that the triad $\{{\bf e}_1,{\bf e}_2,{\bf n}\}$ is not a coordinate basis of $\mathbb{R}^3$, so the components of the dual basis are not closed. This means that we also need the following formulae,
\begin{align}
   d e^{i} &= b^{i}{}_{j}e^j\wedge{\bf n}^\flat\text{,}\\
   d {\bf n}^\flat &= 0\text{,}
\end{align}
as the second fundamental form measures the deviation from a closed coordinate frame of the embedding space. We now give the formulae for partial, covariant and Lie derivatives of the surface vector ${\bf u} = u^{i} {\bf e}_i + u^n {\bf n}$.

The partial derivative is,
\begin{equation}\label{eq:partial_Christoffel}
    \partial_t {\bf u} = \left(\partial_t u^i - u^j v^n b_{j}{}^{i}- u^n\bar\nabla^i v^n\right){\bf e}_i + \left( \partial_t u^n + u^i\bar\nabla_i v^n \right){\bf n},
\end{equation}
and the covariant derivative along the velocity ${\bf v} = v^i {\bf e}_i + v^n {\bf n}$ is given by,
\begin{equation}\label{eq:covariant_Christoffel}
    {\bf v}({\bf u}) = v^i\partial_i\left( u^j {\bf e}_j + u^n{\bf n}\right) = \left(v^i\bar\nabla_i u^j - v^i b_{i}{}^{j} u^n\right){\bf e}_j + \left(v^i u^j b_{ij} +v^i \bar\nabla_i u^n\right){\bf n}\text{.}
\end{equation}
Summing Eq.s~(\ref{eq:partial_Christoffel}) \& (\ref{eq:covariant_Christoffel}) gives the main result (\ref{eq:covariant}). To obtain the exact formula (\ref{eq:covariant}) it is also necessary to replace the surface connections $\bar{\nabla}$ with the ambient space connection $\nabla$, which accounts for the presence of the connection form in (\ref{eq:covariant}). 

The Lie derivative of the vector ${\bf u}$ along the flow ${\bf v}$ is given by,
\begin{align}
    L_{\bf v}{\bf u} &= {\bf v}({\bf u}) -{\bf u}({\bf v}), \\
    &= v^i\partial_i\left( u^j {\bf e}_j + u^n{\bf n}\right) - u^i\partial_i\left( v^j {\bf e}_j + v^n{\bf n}\right)\nonumber\\
    & = \left[ v^i\bar\nabla_i u^j - u^i\bar\nabla_i v^j + b_{i}{}^{j}\left( u^i v^n- v^i u^n\right) \right]{\bf e}_j + \left[ v^i \bar\nabla_i u^n - u^i\bar \nabla_i v^n\right] {\bf n}\text{.}
\end{align}
The computation of the Lie derivative of the $1$-form ${\bf u}^\flat= u_i e^i + u^n {\bf n}^\flat$ is a little more involved. We make use of Cartan's formula $L_{\bf v}{\bf u}^\flat = d \iota_{{\bf v}} {\bf u}^\flat + \iota_{\bf v} d{\bf u}^\flat$. The exterior derivative of ${\bf u}^\flat$ is,
\begin{equation}
    d{\bf u}^\flat = \partial_j u_i e^j\wedge e^i + u_i b^{i}{}_{j}e^j\wedge {\bf n}^\flat + \partial_j u^n e^j\wedge {\bf n}^\flat\text{.}
\end{equation}
We then find that,
\begin{equation}
    L_{\bf v}{\bf u}^\flat = \left[ u_j \bar\nabla_i v^j + u^n \bar\nabla_i v^n + v^j\bar\nabla_j u_i - u_j b^j{}_{i} v^n \right]e^i +\left[ u^i v^j b_{ij} + v^i\bar\nabla_i u^n \right] {\bf n}^\flat\text{.}
\end{equation}
The Lie derivative part of the Jaumann rate is therefore,
\begin{equation}
    \begin{aligned}\label{eq:Lie_sum_Christoffel}
    \frac{1}{2}\left[L_{\bf v}{\bf u} + \left(L_{\bf v}{\bf u}^\flat\right)^\sharp \right] &= \left[ \frac{1}{2} u_j \bar\nabla^i v^j + \frac{1}{2} u^n \bar\nabla^i v^n + v^j\bar\nabla_j u_i - \frac{1}{2} u_j b^{ji}v^n - \frac{1}{2} u^j\bar\nabla_j v^i \right]e_i \\
    & \ \ \ \  +\left[\frac{1}{2} u^i v^j b_{ij} + v^i\bar\nabla_i u^n - \frac{1}{2}u^i\bar\nabla_i v^n\right] {\bf n}\text{.}
    \end{aligned}
\end{equation}
Summing Eq.s~(\ref{eq:partial_Christoffel}) and (\ref{eq:Lie_sum_Christoffel}) gives the Jaumann rate given in (\ref{eq:jaumann}). In this formula the extra terms that appear when replacing surface derivatives by ambient space derivatives cancel. 

\section{Pushforward and Pullback}\label{appC}
Let $f : A \to B$ be a diffeomorphism between manifolds $A, B$. Recall that $f$ induces a linear map $\text{T}f : \text{T}A \to \text{T}B$. In local coordinates $x_i$ on $A$ and $X_j$ on $B$ we may write $f$ in terms of its components $f^j$ as $X^j = f_j(x_1, x_2, \dots,)$, and then the tangent map $\text{T}f$ is just the matrix of partial derivatives $\partial f^j / \partial x_i$. It also defines a dual linear map $\text{T}f^* : \text{T}^*B \to \text{T}^*A$ (the adjoint or transpose of $\text{T}f$), a pullback $f^*$, and a pushforward $f_*$. We give the action of the pushforward and pullback on general tensors. 

On functions:
\\
\adjustbox{scale=1.2,center}{
\begin{tikzcd}
A \arrow{r}{f} \arrow{dr}{g}& B \arrow{d}{f_* g = g \circ f^{-1}}\\
& \mathbb{R}
\end{tikzcd} \ \ \ \ \
\begin{tikzcd}
A \arrow{r}{f} \arrow[swap]{dr}{f^*h = h \circ f} & B \arrow{d}{h}\\
& \mathbb{R}
\end{tikzcd}
}
\\
On vector fields:
\\
\adjustbox{scale=1.2,center}{
\begin{tikzcd}
\text{T}A \arrow{r}{\text{T}f} \arrow{d}{\pi} & \text{T}B \arrow[swap]{d}{\pi} \\%
A \arrow{r}{f} \arrow[u, bend left, "{\bf u}"] & B \arrow[u, bend right, "{\bf w}", swap]
\end{tikzcd}
}
\\
\begin{equation}
    f^* {\bf w} = \text{T}f^{-1} \circ {\bf w} \circ f, \ \ \ \ \ \ f_* {\bf u} = \text{T}f \circ {\bf u} \circ f^{-1}
\end{equation}
On covector fields (1-forms):
\\
\adjustbox{scale=1.2,center}{
\begin{tikzcd}
\text{T}^*A \arrow{d}{\pi} & \text{T}^*B \arrow{l}{\text{T}^*f} \arrow[swap]{d}{\pi} \\%
A \arrow{r}{f} \arrow[u, bend left, "\alpha"] & B \arrow[u, bend right, "\beta", swap]
\end{tikzcd}
}
\\
\begin{equation}
    f^* \beta = \text{T}^*f \circ \beta \circ f, \ \ \ \ \ \ f_* \alpha = (\text{T}^*f)^{-1} \circ \alpha \circ f^{-1}
\end{equation}
On general differential forms and tensors the action is determined by distributivity over the wedge product of forms and the tensor product of tensors. Also note that $f^* = f^{-1}_*$ and vice-versa, so these operations are inverse to one another. 

For a time-dependent family of maps $f_t: A \to B$ we also make use of the time derivatives of the pushforward and pullback of quantities along this map. Let ${\bf w}_t = \partial_t f_t \circ f_t^{-1}$ be the drive velocity, a vector field on $B$. Given a time-independent tensor field ${\bm \sigma}$ on $A$, we have 
\begin{equation}
    \partial_t f^*_t {\bm \sigma} = f^*_t L_{\bf w} {\bm \sigma}, \ \ \ \ \ \ \partial_t f_{t*} {\bm \sigma} = -L_{\bf w} f_{t*} {\bm \sigma}.
\end{equation}
If ${\bm \sigma}_t$ also depends on time, then 
\begin{equation}
    \partial_t f^*_t {\bm \sigma}_t = f^*_t\left(\partial_t{\bm \sigma}_t + L_{\bf w} {\bm \sigma}_t\right), \ \ \ \ \ \ \partial_t f_{t*} {\bm \sigma}_t = f_{t*} \partial_t {\bm \sigma}_t -L_{\bf w} f_{t*} {\bm \sigma}_t. 
\end{equation}

\bibliographystyle{jfm} 
\bibliography{ref}

\end{document}